\documentclass[preprint]{aastex}

\usepackage{graphicx}
\usepackage{rotating}
\usepackage{enumitem}

\newcommand{\Rs}{$ R_{\odot}$}
\newcommand{\msb}{$ B_{msb}$}
\newcommand{\de}{$^{\circ}$}

\newcommand{\pB}{\textit{pB}}

\newcommand{\Btot}{$B_{t}$}
\newcommand{\Bk}{$B_{k}$}
\newcommand{\Bf}{$B_{f}$}

\newcommand{\vcm}{cm$^{-3}$}
\newcommand{\paperii}{Paper \Rmnum{2}}
\newcommand{\paperiii}{Paper \Rmnum{3}}
\newcommand{\iii}{\Rmnum{3}}
\makeatletter

\newcommand{\Rmnum}[1]{\expandafter\@slowromancap\romannumeral #1@}
\makeatother

\begin{document}

\shorttitle{An atlas of coronal electron density \Rmnum{1}}
\shortauthors{Morgan}
\title{An atlas of coronal electron density at 5\Rs \\
\Rmnum{1}: Data processing and calibration}
\author{Huw Morgan} 
\affil{Institute of Mathematics, Physics and Computer Science, Aberystwyth University, Ceredigion, Cymru, SY23 3BZ}
\email{hmorgan@aber.ac.uk}

\begin{abstract}
Tomography of the solar corona can provide cruicial constraints for models of the low corona, unique information on changes in coronal structure and rotation rates, and a valuable boundary condition for models of the heliospheric solar wind. This is the first of a series of three papers which aim to create a set of maps of the coronal density over an extended period (1996-present). The papers will describe the data processing and calibration (this paper), the tomography method (\paperii) and resulting atlas of coronal electron density at a height of 5\Rs\ between years 1996-2014 (\paperiii). This first paper presents a detailed description of data processing and calibration for the Large-Angle and Spectrometric Coronagraph (LASCO) C2 instrument onboard the Solar and Heliospheric Observatory (SOHO) and the COR2 instruments of the Sun Earth Connection Coronal and Heliospheric Investigation (SECCHI) package aboard the Solar Terrestial Relations Observatory (STEREO) A \& B spacecraft. The methodology includes noise suppression, background subtraction, separation of large dynamic events, conversion of total brightness to K-coronal brightness and simple functions for crosscalibration between C2/LASCO and COR2/SECCHI. Comparison of the brightness of stars between LASCO C2 total and polarized brightness (\pB) observations provide in-flight calibration factors for the \pB\ observations, resulting in considerable improved agreement between C2 and COR2 A, and elimination of curious artifacts in the C2 \pB\ images. The crosscalibration between LASCO C2 and the STEREO coronagraphs allows, for the first time, the potential use of multi-spacecraft coronagraph data for tomography and for CME analysis. 
\end{abstract}

\maketitle

\section{Introduction}

Historically, measurements of the coronal white light brightness were restricted to total solar eclipse observations. Polarized brightness measurements, which are dominated by Thomson-scattered light from coronal electrons, allowed the inversion of observations to estimate coronal electron densities \citep{van1950}. Such inversion methods assume a simple geometry (spherical or cylindrical) to the distribution of density, and result in a useful estimate of the overall drop in density with height within streamers or coronal holes \citep[e.g.,][]{van1950,saito1977,quemerais2002}. They are most accurate during solar minimum where the distribution of density is closer to the assumed geometry \citep{guhathakurta1996,gibson2003, morgan2007fcorona}. 

The rapid development of instrumentation over the past few decades, in particular space-based coronagraphs, allow regular high-cadence observations. This allows coronal rotational tomography methods to be applied, resulting in estimates of the coronal density distribution without the need to assume a geometry. That is, the line of sight may be resolved. A successful tomography method was devised by \citet{frazin2000}, enabling reconstructions even during solar maximum \citep{butala2005}. These techniques aim to find a distribution of electron density in a 3D corona which best satisfy a set of coronagraphic \pB\ observations made over half a solar rotation (half a rotation since both east and west limbs are observed), subject to some reasonable assumptions such as the smoothness of the reconstruction. These techniques have been applied to very low heights in the corona \citep{kramar2014}, which bodes well for better constraints on large-scale magnetohydrodynamic (MHD) models at these heights. A novel method for creating qualitative maps of the distribution of coronal structure was introduced by \citet{morgan2009}, resulting in a comprehensive study of coronal structure over a solar cycle \citep{morgan2010structure} and measurements of coronal rotation rates \citep{morgan2011rotation}.

Regardless of the tomography method, the accuracy of the reconstructed densities are dependent on both the accuracy of the input data (calibration and possibly background subtraction uncertainties) and the amount of available data (most tomographical methods are limited to using polarized brightness observations in order to avoid the need for background subtraction). It is desirable to use the large archive of total brightness observations for tomography, rather than the far lower number of polarized brightness observations. \citet{frazin2010} show one method to achieve this. 

Another major challenge is dealing with changes in the coronal structure over the observational period (half a solar rotation). Currently, with three operational spacecraft observing the corona from different directions, it is in principle possible to reduce the time period needed for a tomographical reconstruction by using the multi-instrument data as input for tomography, but this depends on accurate crosscalibration. A recent work to discuss calibration and crosscalibration of the available instruments is \citet{frazin2012}. Both \citet{frazin2010} and \citet{frazin2012} have considerable relevance to this work. Absolute calibration is another issue. Even with good cross-calibration between coronagraphs, a decision must be made as to which instrument is used as a standard. Most works depend on the standard calibration procedures provided by the instrument groups, generally based on pre-flight laboratory measurements \citep[e.g.][]{brueckner1995,howard2002}. The most advanced absolute calibration is gained by the use of stars observed within the coronagraph fields of view, giving the most precise in-flight calibration over long time periods \citep[][and references within]{gardes2013,colaninno2015}.

Tomographical maps of the coronal density have many useful applications, including coronal rotation rates \citep{morgan2011rotation}, relationships between large-scale coronal structure and solar features \citep{morgan2010structure,morgan2011solmin}, constraints or comparison with magnetic or MHD models \citep{morgan2009,kramar2014}, for interpreting observations of other instruments \citep{frazin2003}, and for extrapolation into the heliosphere. Large-scale models of the corona and heliosphere would benefit greatly from an additional empirical constraint of density distribution in the extended corona.

Reliable crosscalibration of coronagraphs is useful for purposes other than coronal tomography, most notably for coronal mass ejection (CME) analysis. Estimates of CME masses contain large uncertainties due to lack of information on their true 3-dimensional distribution, and it is only through correct crosscalibration of coronagraphs viewing from different directions that this may be resolved \citep{frazin2009,frazin2012cme}. Also useful for this type of analysis are methods for isolating the CME from the background coronal structures \citep{morgan2010cme,morgan2012cme}.

A brief introduction to the instruments and observations is in section \ref{inst}. The technique to separate the dynamic and quiescent coronal components is described in section \ref{dst}. Calibration of LASCO C2, including background subtraction, is in section \ref{c2}, 
followed by crosscalibration of LASCO C2 and COR2 A (section \ref{c2cora}) and of COR2 A \& B (section \ref{cor2ab}). An overview of how the calibrated data may be utilised in future work is in section \ref{application}, and section \ref{summary} gives a summary.

\section{Instruments and observations}
\label{inst}
\subsection{LASCO C2}

The Large Angle and Spectrometric Coronagraph (LASCO, \citet{brueckner1995}) C2 instrument aboard the Solar and Heliospheric Observatory (SOHO, \citet{domingo1995}) has observed the extended inner corona almost continuously since 1996. It has an useful field of view of $\sim$2.2 to 6.0\Rs\ (measured from solar disk center), collected on a 1024 by 1024 detector (with occasional onboard rebinning to 512 by 512). The cleanliness and excellent quality of the data over almost two decades of operation is a testament to the instrument builders and team. LASCO C2 can use several different wide bandpass filters. This study uses the orange filter exclusively, being by far the most widely used observing filter. The data is publically available from many sources in the form of standard fits files. These level 0.5 fits files are images containing the detector counts plus a header which records all the essential observational information (pointing, dates, exposure times etc.). There is a large set of software to open and process the LASCO fits files in the Solarsoft library.

LASCO C2 makes two types of observations - total brightness (\Btot) and polarized brightness (\pB). The \pB\ observations are made typically once or twice a day, and involve a sequence of observations made using polarizers at varying angles. These observations may be combined during calibration to form an estimate of the coronal \pB. The total brightness observations are made far more frequently, varying from $\sim$30 minutes at the start of the mission, up to $\sim$10 minutes by 2014. Typical exposure times are $\sim$25s.

LASCO C2 \Btot\ images contain three main components: instrumental stray light, the K corona brightness (\Bk), and the F corona brightness (\Bf), with \Bf\ becoming dominant at heights above $\sim$3\Rs\ \citep[e.g.,][]{van1950}. \Bf\ is the emission of scattered sunlight from interplanetary dust integrated along a line of sight, first described by \citet{grotrian1934}. \Bk\ is Thomson-scattered emission from coronal electrons, also integrated along a line of sight \citep{van1950}. Isolating the F and K components of brightness from C2 observations is not a trivial matter. 

Stray light and \Bf\ is largely unpolarized below 6.0\Rs\ \citep{saito1977,mann1992,morgan2007fcorona}. The \pB\ images are therefore mostly free of \Bf. This is exploited in this work as a means to create background images of the F-corona and stray light which may be subtracted from the \Btot\ to isolate \Bk. A proper calibration of the \pB\ images is important. To achieve this, we use a comparison of star brightness between the \pB\ and \Btot\ images to obtain in-flight values of corrected calibration factors for the \pB\ sequences. The \pB\ images can subsequently be radiometrically calibrated using published values based of \Btot\ calibration factors, also gained from stars \citep{colaninno2015}.

\subsection{SECCHI COR A \& B}
The COR2 coronagraphs, part of the Sun Earth Connection Coronal and Heliospheric Investigation (SECCHI, \citet{howard2002}) suite of instruments aboard the twin Solar Terrestial Relations Observatory (STEREO A \& B, \citet{kaiser2005}), have been observing the corona since 2005. They have useful fields of view from $\sim$4-14\Rs\ collected on a 2048 by 2048 detector. As with LASCO, there is a large set of software to open and process the COR2 fits files in the Solarsoft library.

COR2 makes both total brightness (\Btot) and polarized brightness (\pB) sequence observations. In contrast to LASCO C2, the \pB\ observations are made far more frequently - typically every half hour or so. In principle, the LASCO C2 procedures to create calibrated backgrounds for subtraction from total brightness observations (described later) could also be applied to COR2 total brightness observations. In practice, there is not much to be gained in doing this - \pB\ observations every half-hour is sufficient time resolution for effective tomography and also for CME analysis, without the complication and uncertainty of additional calibration steps. The standard calibrations (Solarsoft routine secchi\_prep.pro) are applied to the COR2 \pB\ sequences to obtain \pB\ images. These are then transformed into an approximation of \Bk\ using an inversion/integration procedure similar in concept to \citet{saito1977}, \citet{quemerais2002} or \citet{morgan2007fcorona} to be described in detail later.

\section{Dynamic separation technique}
\label{dst}
The dynamic separation technique (DST) was introduced in concept by \citet{morgan2010cme} and further developed into a spatio-temporal deconvolution method by \citet{morgan2012cme} in the context of automated CME detection \citep{byrne2012}. This section describes an improved method which can be applied to coronagraph data without the steps of background removal and normalizing-radial-graded-filter (NRGF) which were used previously. As such, it is more robust and also serves to remove certain static instrumental errors from the data. The main purpose of the DST is to split the coronagraph images into two components - the static, quiescent corona (including F-corona and instrumental background) and dynamic events. Successful separation of the two components is a very powerful tool for study of both the quiescent corona and CMEs or smaller dynamical features \citep{morgan2013expandingloops1,morgan2013expandingloops2}. It also aids in creating stable background images for subtraction, and is important in achieving reliable calibration as will be shown in following sections.

\begin{figure}[]
\begin{center}
\includegraphics[width=10.0cm]{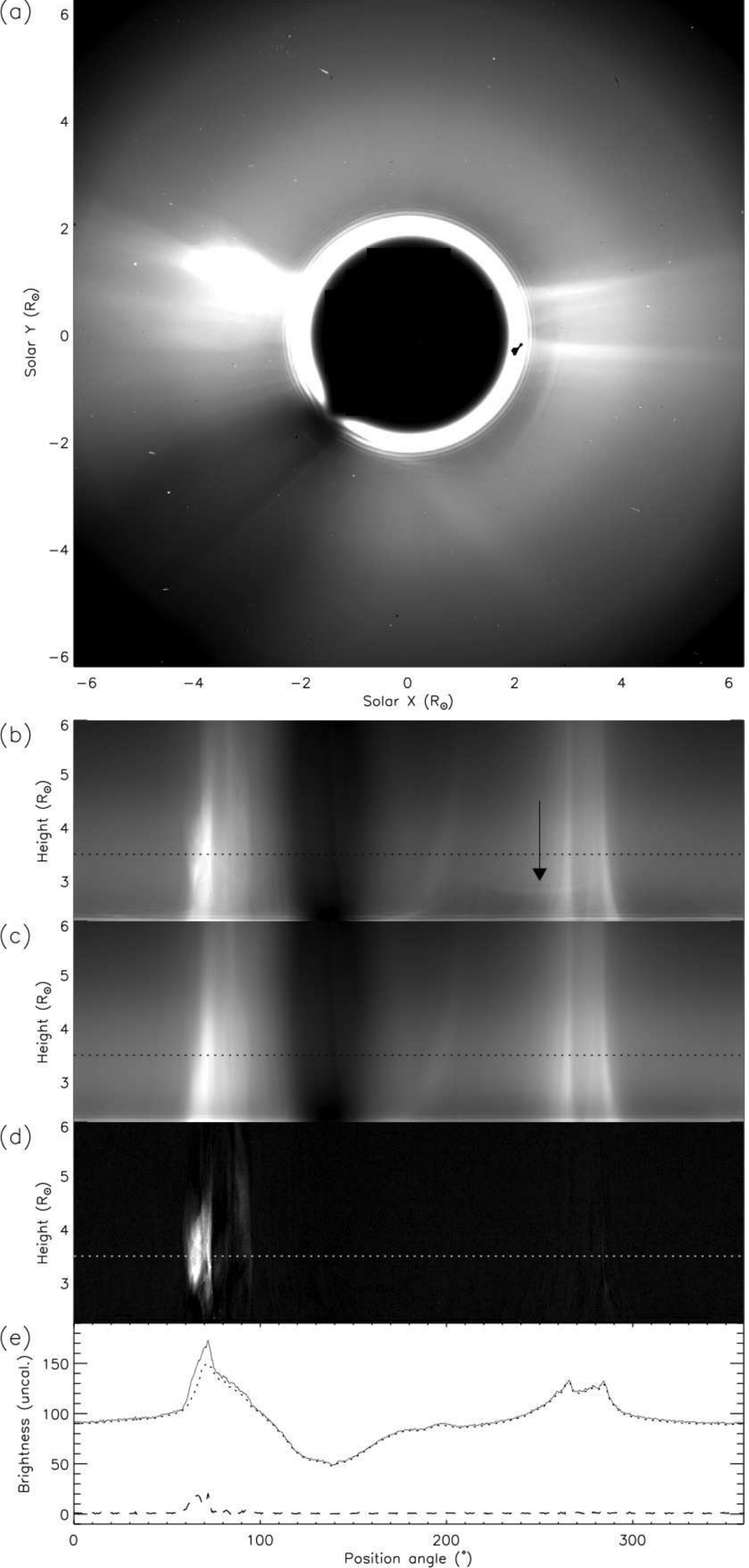}
\end{center}
\caption{LASCO C2 observation of 2007/03/21 11:26 at various steps of the separation method: (a) Original image following unwarping and exposure time normalization. (b) The same image following a point filter to remove isolated pixels with spurious values, and the field of view from heights 2.25 to 6.00 transformed into polar coordinates (position angles measured counter-clockwise from north). The arrow points to an instrumental stray light feature. (c) The quiescent component of the polar image following the DST. (d) The dynamic component. (e) One `slice' of the original (solid), quiescent (dotted) and dynamic (dashed) images showing uncalibrated brightness at a height of 3.5\Rs. This height is indicated by the dotted lines in (b)-(d).}
\label{figsep}
\end{figure}

Figure \ref{figsep}a shows a LASCO C2 total brightness observation of 2007/03/12 11:26, which includes a small CME in the North-east corona. The image is a standard level 0.5 fits file, normalized by the exposure time and with the image unwarped or distorted to provide the correct observational geometry using the standard LASCO procedure c2\_warp.pro. The image is cleaned using a point filter. The filter iteratively identifies isolated pixels, or small group of pixels, with very large or small brightness compared to the mean and standard deviation of their local region, and replaces their values with the local mean with each iteration. The image is transformed into polar coordinates, limiting the field of view (FOV) to heliocentric heights of 2.25-6.00\Rs. This polar image is shown in figure \ref{figsep}b, with 720 position angle bins and 300 height bins. The rest of the processing and calibration is applied in polar coordinates - it is convenient to work in this coordinate space, particularly for the DST. The arrow in figure \ref{figsep}b point to a large instrumental stray light artifact which will be effectively removed by the DST.

\begin{figure*}[]
\begin{center}
\includegraphics[width=12.5cm]{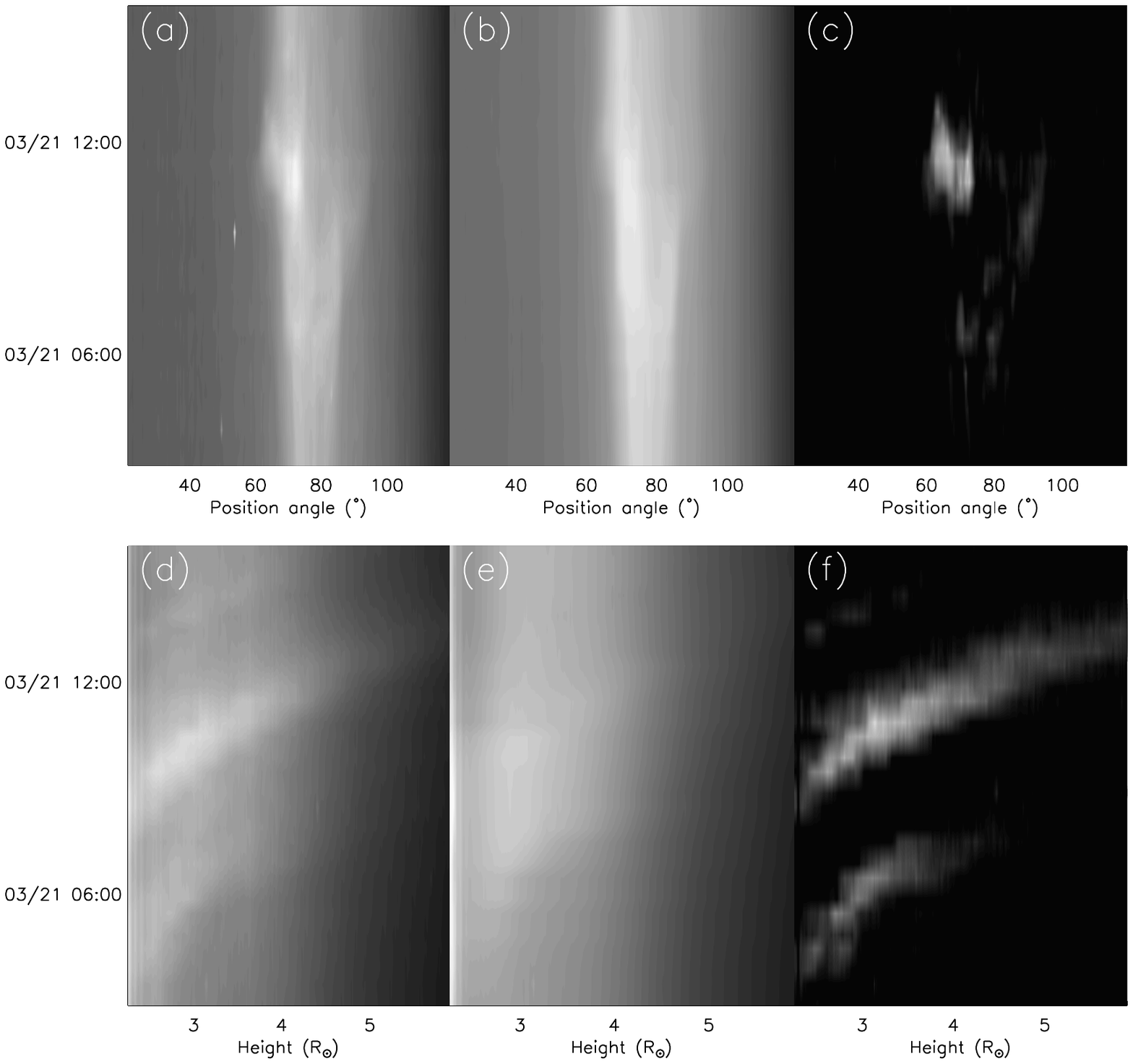}
\end{center}
\caption{(a)-(c) Angle-time and (d)-(f) height-time stack plots showing original, quiescent and dynamic component respectively for LASCO C2 observations during 2007/03/21.}
\label{figsep2}
\end{figure*}

The DST works under the assumption that the quiescent corona in the LASCO C2 field of view is close to radial (at least smooth in the radial direction), and changes only slowly (or smoothly) in time. Dynamic events form localised regions which are not smooth in the radial, and by definition, change rapidly in time. Thus applying an iterative deconvolution in the radial and time dimensions, with appropriate choice of smoothing kernels, leads to an estimate of a smooth background (quiescent component) with the residual forming an estimate of dynamic events (dynamic component). A box-car kernel of width 0.3\Rs\ in the radial and up to 8 hours in time is used here. This corresponds to around 24 radial bins and around 45-55 observations depending on cadence. Irregular cadence is not a large problem for the method, providing that there are a reasonable number of images in the 8 hour window. Therefore, for our example image of 2007/03/21 11:26, there are 50 total brightness observations made between 08:26 and 15:26 ($\pm4$ hrs). These are all transformed to the same polar coordinates, forming a datacube in position angle, radial dimension, and time. An example of a height-time slice, ready for DST processing at position angle 70\de\ is shown in figure \ref{figsep2}c. The small CME can be seen as two brighter streaks against the background. 

The deconvolution is an iterative process. The first step is a temporary reduction in the radial drop-off in brightness. Over the whole data cube (all position angles and times), $m_r$ ,the median decrease of brightness against height, is calculated. If $I$ is the original polar image, a new image is given by $I_r=I / m_r$ . Initially, the dynamic component $D_{0}$ is everywhere zero and the initial quiescent component $Q_{0}$ is set equal to $I_r$. At iteration $i$ ($=1,2,3...$) the estimate of the dynamic component $D_{i}$ and the quiescent component $Q_{i}$ are updated by 
\begin{eqnarray}
D_{i} = D_{i-1} + [( Q_{i-1}-Q_{i-1} \otimes k )>0], \\
Q_{i} = I_r - D_{i}, 
\label{eqndst}
\end{eqnarray}
where $k$ is the radial-time boxcar kernel. Iteration ends when the total absolute difference between subsequent $D_i$ drops below a certain threshold or when $i=10$. The dynamic and quiescent components are then multiplied along the radial dimension by $m_r$, thus reintroducing the radial drop-off in brightness. This is the main separation routine, and the resulting quiescent images are very clean as shown in figure \ref{figsep}c. These seem to be free of any streaks or diffraction patterns near the inner FOV. The instrumental error indicated by an arrow in figure \ref{figsep}b is not present. Diffraction patterns and other instrumental artifacts are present in the dynamic component. These are easily removed by calculating for each position angle and height bin within the dynamic images the median value over a time range of $\sim\pm8$ hours from the time of interest, and subtracting from the current dynamic image. The result is a set of very clean images containing real dynamic events and no artifacts, as shown in figure \ref{figsep}d. Figure \ref{figsep}e shows constant-height cuts across the original, quiescent and dynamic images at a height of 3.5\Rs. 

Whilst figures \ref{figsep}b-d show the results of the separation process for a single observation, figure \ref{figsep2} shows angle-time  and height-time stacks for the CME event of 2007/03/21. These figures confirm the cleanliness of the DST and the effective removal of instrumental artifacts. Figure \ref{figsep2}e shows the height-time quiescent component. There is a hint of enhanced brightness at the position of the CME in this image which shows that the separation of dynamic and quiescent cannot be made perfectly because structures in the quiescent corona move in response to the CME.  A critical choice therefore is the choice of kernel width. A narrower kernel (in radial and time dimensions) would enhance smaller-scale features in the dynamic images whilst allowing more of the larger-scale dynamic components into the quiescent images. The choice of kernel presented here is gained from trial and error for many types of CMEs. What is important is that most of the CME signal is found in the dynamic component whilst we cannot avoid a small, smooth change in the quiescent component during the passage of larger CMEs which do cause a change in the distribution of quiescent coronal structure. 

\begin{figure*}[]
\begin{center}
\includegraphics[width=12.5cm]{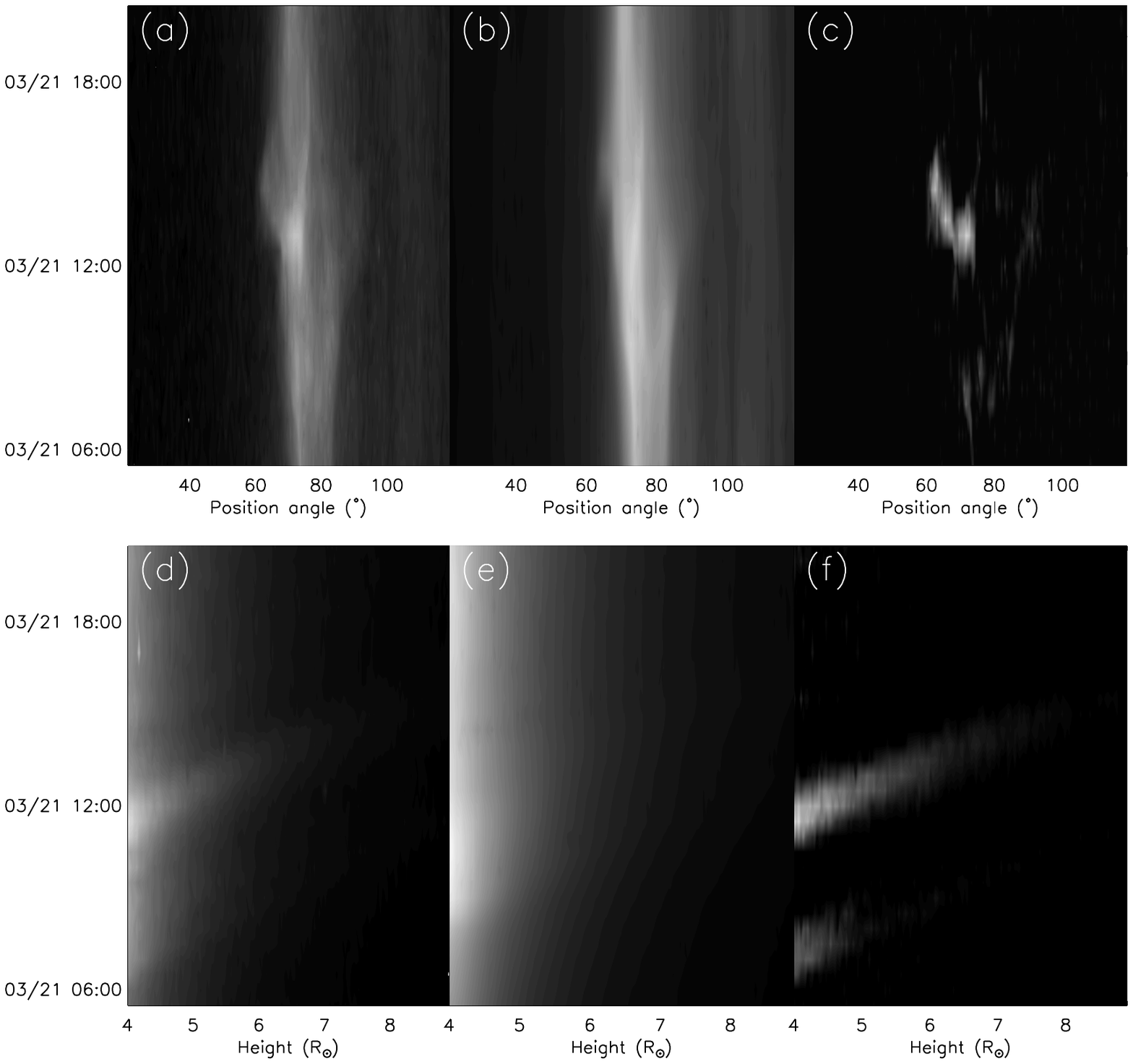}
\end{center}
\caption{(a)-(c) Angle-time and (d)-(f) height-time stack plots showing original, quiescent and dynamic component respectively for SECCHI COR2 A observations during 2007/03/21.}
\label{figsep3}
\end{figure*}

An identical DST method is applied to the SECCHI COR2 coronagraphs, with similar results. Figure \ref{figsep3} shows angle-time and height-time stack plots for the same CME event as shown for LASCO C2 in figure \ref{figsep2}, but for SECCHI COR2 A. At this time, SOHO and STEREO A are close, therefore the CME should, and does, appear structurally similar. What is considerably different, of course, is the actual brightness values. Use of the COR2 polarized brightness sequences allows the separation to be made on the COR2 calibrated data, in units of mean solar brightness (\msb). At this stage of processing, the LASCO C2 data are not yet calibrated. Comparison of actual brightness values and estimated CME mass is made later in section \ref{application}.

\section{Initial calibration of LASCO C2}
\label{c2}
Except for a few dedicated observational campaigns, \pB\ measurements are made infrequently by LASCO C2. Typically one or two measurements are made per day. Total brightness (without the polarizers) are made very often, with a cadence of $\sim$20 minutes, increasing to $\sim$10 minutes in more recent years (due to an increasing portion of SOHO's telemetry as other instruments become non-operational). For tomography of the corona, and for properly calibrated measurements of dynamic events, it is important to properly calibrate the total brightness observations. 

In this section, the infrequent \pB\ observations are compared to the total brightness observations and are used as `tie-points' to form calibrated backgrounds which may be subtracted from the total brightness observations. The main steps are:
\begin{itemize}
\item Converting the \pB\ observations into \Bk\ values by inversion.
\item Comparing the \Bk\ images to radiometrically calibrated total brightness images (the quiescent images gained from the DST) to form one background image per \pB\ observation.
\item Creating a final long time-period ($\sim$10 day) median of the individual background images.
\end{itemize}
Thus quiescent component \Btot\ images may be transformed to our approximation of \Bk, by multiplication with a standard radiometric calibration and vignetting image, and subtraction of a long-term background image which contains stray light and \Bf. The resulting calibrated total brightness images are shown to agree well with the \Bk\ values gained from the \pB\ observations, as should be expected given the procedure. Before describing the procedure, the following section uses observations of stars to obtain new calibration factors for the \pB\ observations.

\subsection{Calibration using stars}

The calibration of LASCO C2 total brightness observations has been made using a detailed study of star brightness, providing a calibration factor of low uncertainty which changes linearly over time \citep{colaninno2015}. In this section, the brightness of stars is compared between the \Btot\ and \pB\ observations of LASCO C2, providing a new inflight calibration factor for the \pB\ observations. The stars are also used to test the LASCO C2 \Btot\ flat field, or vignetting, correction. This analysis is based on the assumption that the collected star light is largely unpolarized, which is a valid assumption based on works of stellar astronomy (see table 1 of \citet{fosalba2002}, and references within their introduction).

\subsubsection{Automated identification and tracking of stars}
\begin{figure}[]
\begin{center}
\includegraphics[width=7.0cm]{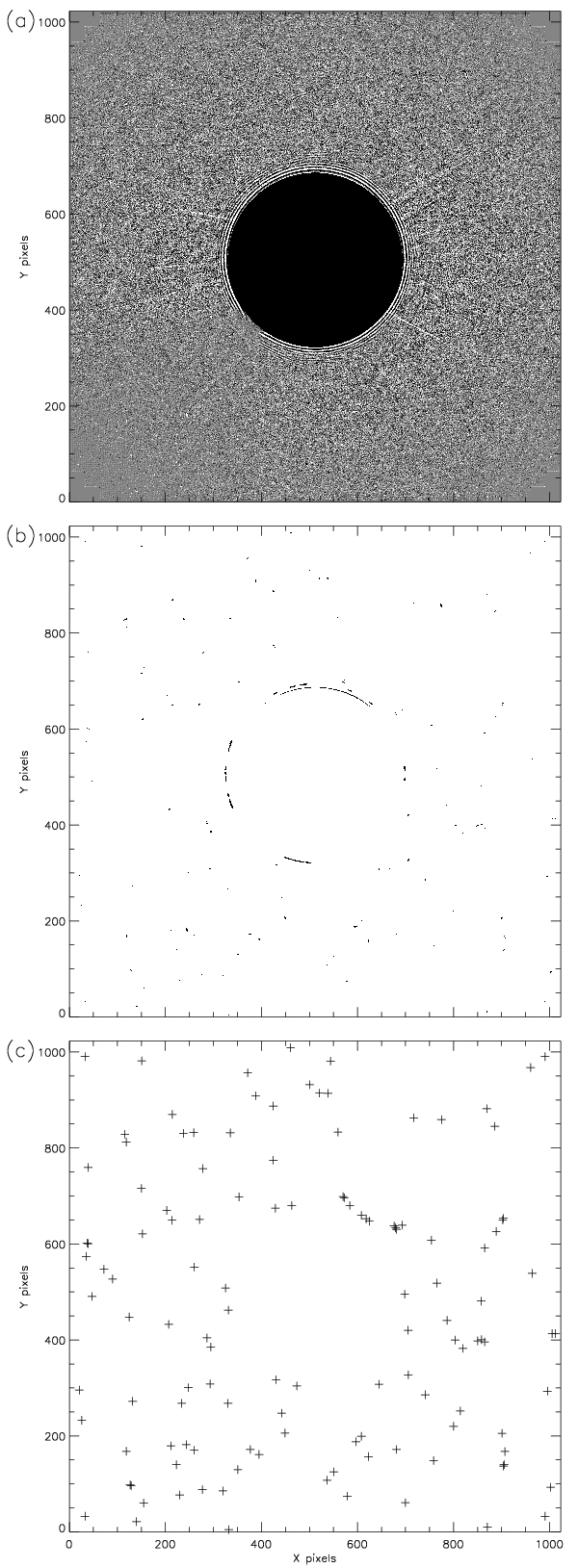}
\end{center}
\caption{The identification of bright pixels or group of pixels in a LASCO C2 image, as a first step towards tracking stars. (a) A median-filtered image containing bright points and edges of coronal structures. (b) Binary image showing isolated bright pixels or groups of pixels. (c) Points centered on each bright pixel or group of pixels.}
\label{testmedian2}
\end{figure}

\begin{figure}[]
\begin{center}
\includegraphics[width=8.0cm]{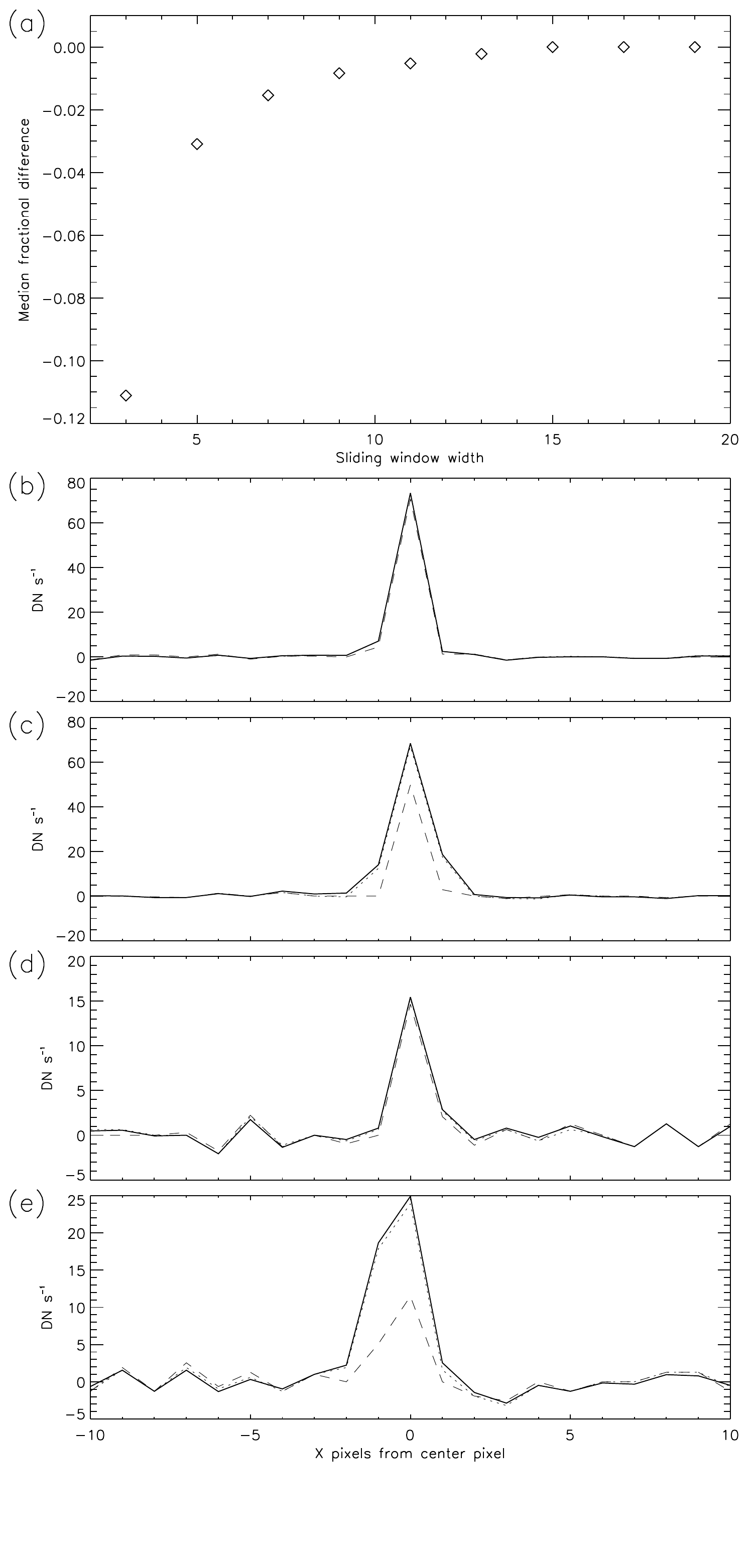}
\end{center}
\caption{Testing the size of the median sliding-window filter. (a) The mean relative brightness of bright points as the size of the sliding window is increased from 3 to 19 squared. This is a mean fractional difference across all bright pixels in the image shown in figure \ref{testmedian}, with the brightness compared to the value obtained for the largest window (19 pixels squared). (b)-(e) Cuts across the median-filtered images through 4 example bright points. The solid line is for the 19-pixel squared window, the dashed line is for the 3-pixel squared window. A dotted line is for a 11-pixel squared window, hidden behind the solid line due to having close to identical brightness values for most points in the image.}
\label{testmedian}
\end{figure}

From year 2000 to 2015, all LASCO C2 \Btot\ observations using the orange filter are processed to identify and isolate the signal from bright points. Point-like features are revealed by subtracting a median-filtered image, with sliding window of width $11\times11$ pixels. This is applied to the raw level 0.5 fits files. Such a high-pass image is shown in figure \ref{testmedian2}a. The choice of sliding window size is important. Too small would risk discarding pixels containing star signal. As the sliding window size is increased, then more of the background coronal structures are included. For example, some faint rays at the boundaries of bright streamers can be seen in figure \ref{testmedian2}a. Figure \ref{testmedian} shows how the choice of sliding window size can effect the measured brightness for point-like objects in an image. Figure \ref{testmedian}a shows the mean fractional difference between brightnesses for all point-like objects in a LASCO C2 image as the size of the sliding window is increased from 3 to 19 pixels squared. For the purpose of comparison, the brightness for the largest sliding window is taken as the standard, or target. Our choice of 11 pixels squared is a sensible compromise to avoid `leakage' from background coronal structure, and to preserve the desired brightness. This is confirmed by figures \ref{testmedian}b-e which show cuts across several bright pixels for sliding windows of size 3, 11, and 19 pixels. As expected, there is a large difference between the smallest window and the others, whilst the 11 and 19 pixel-squared windows are almost identical.

The point features are automatically identified in the high-pass image by setting a threshold of $\geq 11$ counts/second, thus creating a binary mask with values of 1 containing the brightest points, above the threshold. Stars may have signal spread over more than one pixel, so that pixels neighbouring the brightest pixel are lower than the threshold. To allow for this, the binary image is convolved with a narrow gaussian kernel, and all pixels with value greater than a very small number in the smoothed binary image, and value $\geq$ 3 times the local median absolute deviation in the high-pass image, are finally identified as pixels which may contain signal from stars. In this way, if pixels contain significant counts, and are close to the brightest pixels, they are included in the final estimate of brightness. The black pixels in figure \ref{testmedian2}b are the pixels identified as potential stars. Many of these pixels are grouped together in small numbers. Finally, each candidate pixel, or group of pixels, is recorded as a single point, as shown in figure \ref{testmedian}c. The brightness of each point is given by the total brightness of all pixels in each group, and its position given by the mean $x$ and $y$ of all pixels in the group, weighted by their brightness. Groups too close to the occulting disk are discarded.

\begin{figure}[]
\begin{center}
\includegraphics[width=14.0cm]{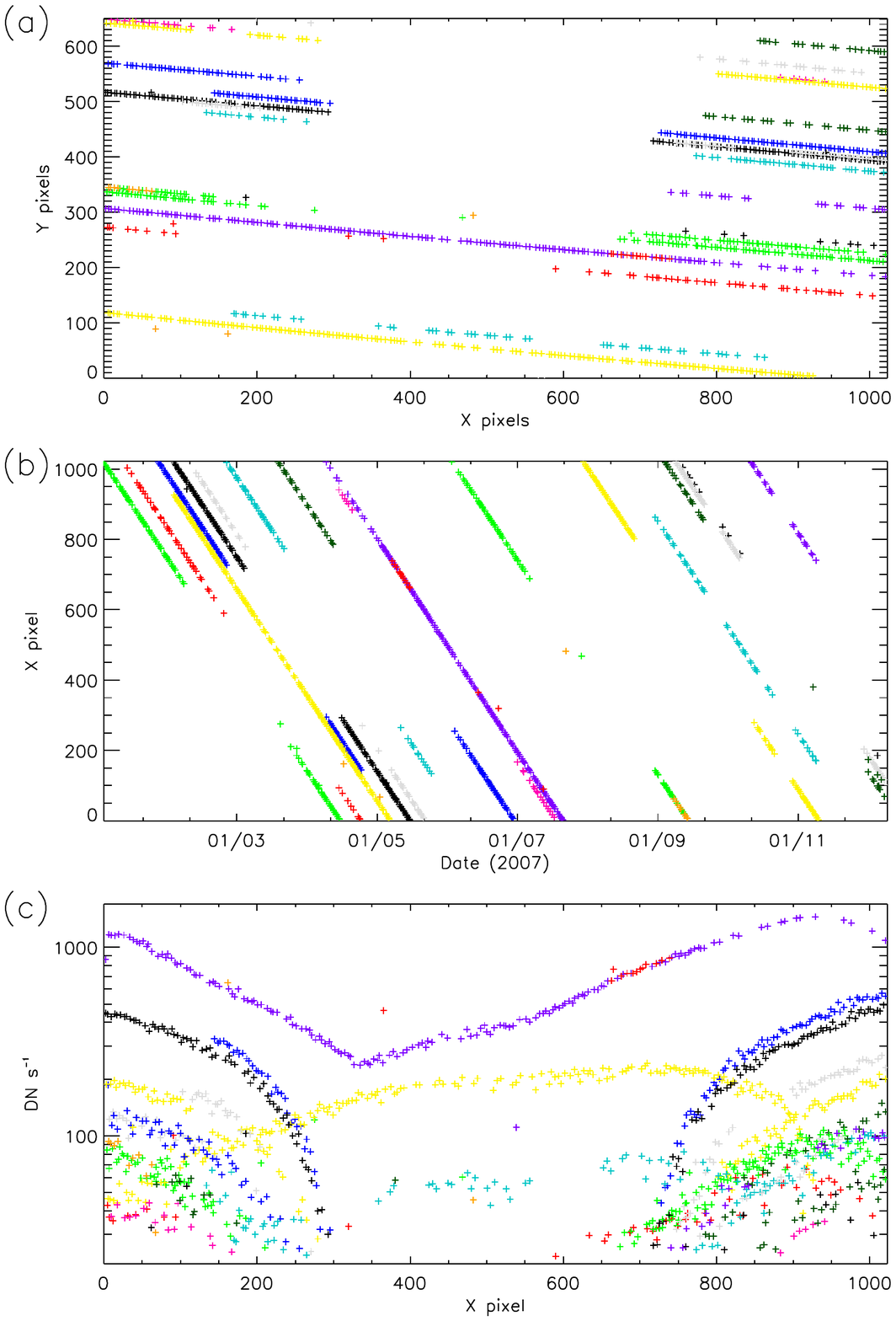}
\end{center}
\caption{Following the identification of star tracks using two Hough transforms in space and time, this figure shows the tracks of stars in (a) image coordinates over time and (b) across the image $x$-coordinate in time. Colours are used to distinguish different star tracks, although note that the use of some colours is repeated for several tracks. (c) shows the time-normalised brightness of these stars as a function of image $x$.  }
\label{startracks}
\end{figure}

The recording of bright pixels is repeated throughout the $\sim$15 years of images. All bright pixels are recorded, stars or otherwise. By use of a Hough transform, first in space coordinates, then in time, all bright points which do not move through images in time as one would expect stars to move, are discarded. So bright pixels describing straight lines in the image $x-y$ coordinates, and also at the required slope in $x-t$ coordinates, and containing a decent amount of points, are identified as stars. The tracks of stars in both space and time are shown in figure \ref{startracks}a and b respectively for a period of two weeks in January 2007. Note that the filtering method is effective at identifying the track of a star even across the occulter, and of grouping points according to star. The time-normalized counts of each star track is shown as a function of image-$X$ in figure \ref{startracks}c. Short periods of time bridging spacecraft rotation manoeuevres force gaps of a few days in the identification and grouping process.

\subsubsection{Testing the LASCO C2 \Btot\ flat field}
The first useful application of the tracks of stars over a long time period is as a check on the LASCO C2 flatfield, or vignetting, correction. This correction is provided in the standard C2 calibration routines within the Solarsoft routines as the fits file \emph{c2vig\_final.fts}. Each star shows a well-behaved variation in brightness as it crosses the C2 field of view, as can be seen in figure \ref{startracks}c. This variation shows the response of the instrument to a constant signal. Each star track's brightness is normalized by its 99 percentile maximum (i.e. robust maximum). These normalized values are then collected for large bins across the whole image. For each bin, a robust mean and standard deviation is calculated. For the robust mean, values larger than twice the median absolute deviation from the median are discarded, and a mean calculated for the remaining values. This results in an estimate of the flat field across the whole image barring heights close to, or within, the occulter. There is a potential flaw in this procedure in the assumption that each star's track maximum coincides with a maximum in the flat field. This assumption is valid for C2 given the circular form of the flat field and that the vast majority of star tracks extend across the whole image, and must therefore pass through some maximum point in the flat field. 

\begin{figure*}[]
\begin{center}
\includegraphics[width=14.0cm]{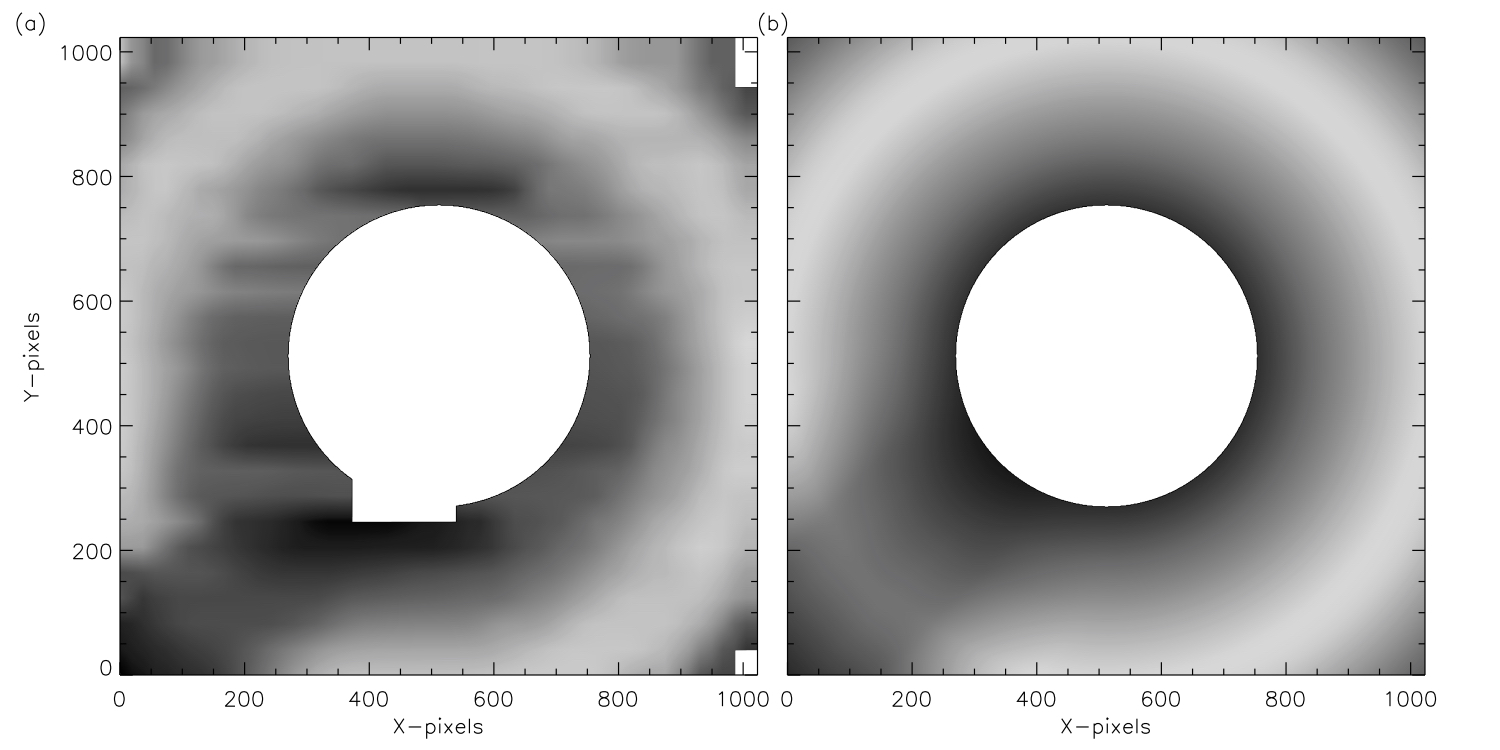}
\end{center}
\caption{(a) Flat field for LASCO C2 calculated from the brightness of stars between years 2000-2015. (b) The standard C2 flat field given by the Solarsoft file \emph{c2vig\_final.fts}.}
\label{flatfield}
\end{figure*}

\begin{figure}[]
\begin{center}
\includegraphics[width=12.0cm]{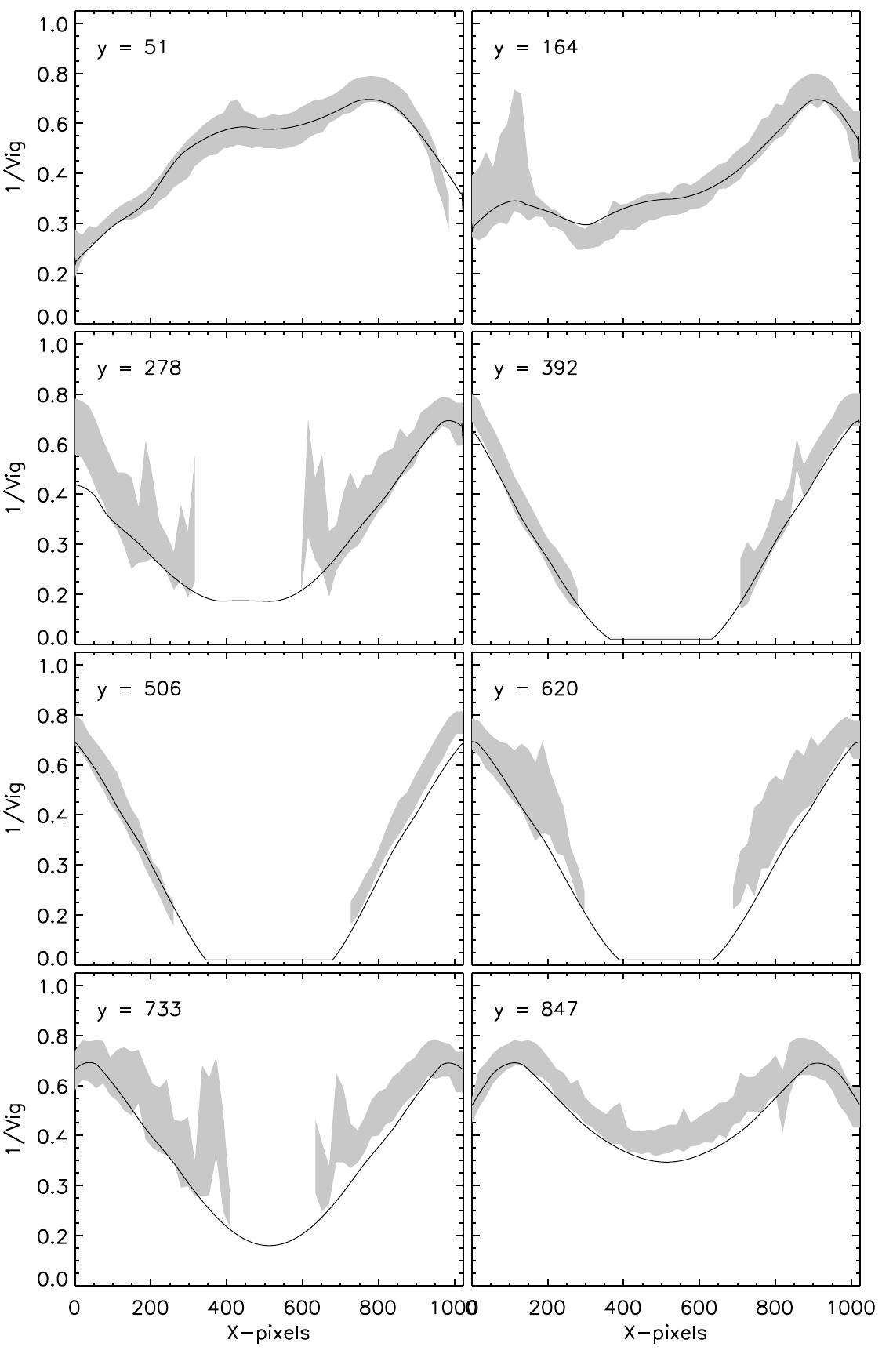}
\end{center}
\caption{Horizontal cuts across the LASCO C2 flat field for the standard Solarsoft C2 flat field (thick solid line), and for the estimated flat field from stars (shaded grey areas). The shaded areas show the robust mean $\pm$ one standard deviation. The $Y$ position of each cut is given in each plot.}
\label{flatfield2}
\end{figure}

The star flat field is compared to the standard C2 correction for the whole field of view in figure \ref{flatfield}, and plotted for a variety of cuts across the image in figure \ref{flatfield2}. The standard flat field falls within the margins of error of the estimated star flat field across most of the image. Small regions of disagreement coincide with regions where the star values have a large variation - often near the occulter. Based on this analysis, the standard C2 flatfield provided by the instrument team in Solarsoft is reliable, and will be used with confidence in the remainder of this work.

\subsubsection{Star calibration of LASCO C2 \pB\ observations}
All \pB\ observation made by LASCO C2 in years 2000-2015 with the orange filter is used in this analysis. Each file is examined for point-like features following the procedures described above for the \Btot\ observations (figure \ref{testmedian2}), but with a suitably low brightness threshold for each polarizer position, and a reduced-size sliding window median filter which is discussed in the following paragraph. The position and observation time of each detected point-like feature is then compared to the tracks of stars identified in the \Btot\ observations. This is achieved by fitting the spatial position of each \Btot\ star track to a straight line. The expected position of each star at the time of the \pB\ observation is then compared to the position of detected points in the \pB\ image. If the distance is less than 1 pixel, the bright \pB\ point is considered a star, and its brightness can be directly compared to the brightness of the star in the \Btot\ observations. To achieve this, the brightness of the \Btot\ star track is fitted to a $4^{th}$ degree polynomial, as a function of the $X$-pixel position across the image. This allows interpolation to the time of the \pB\ observation, and a cleaner estimate of the expected brightness (in comparison to taking the brightness of the nearest neighbour, for example). Finally, the ratio of the \pB\ brightness to the estimated \Btot\ brightness is recorded.

\begin{figure}[]
\begin{center}
\includegraphics[width=7.0cm]{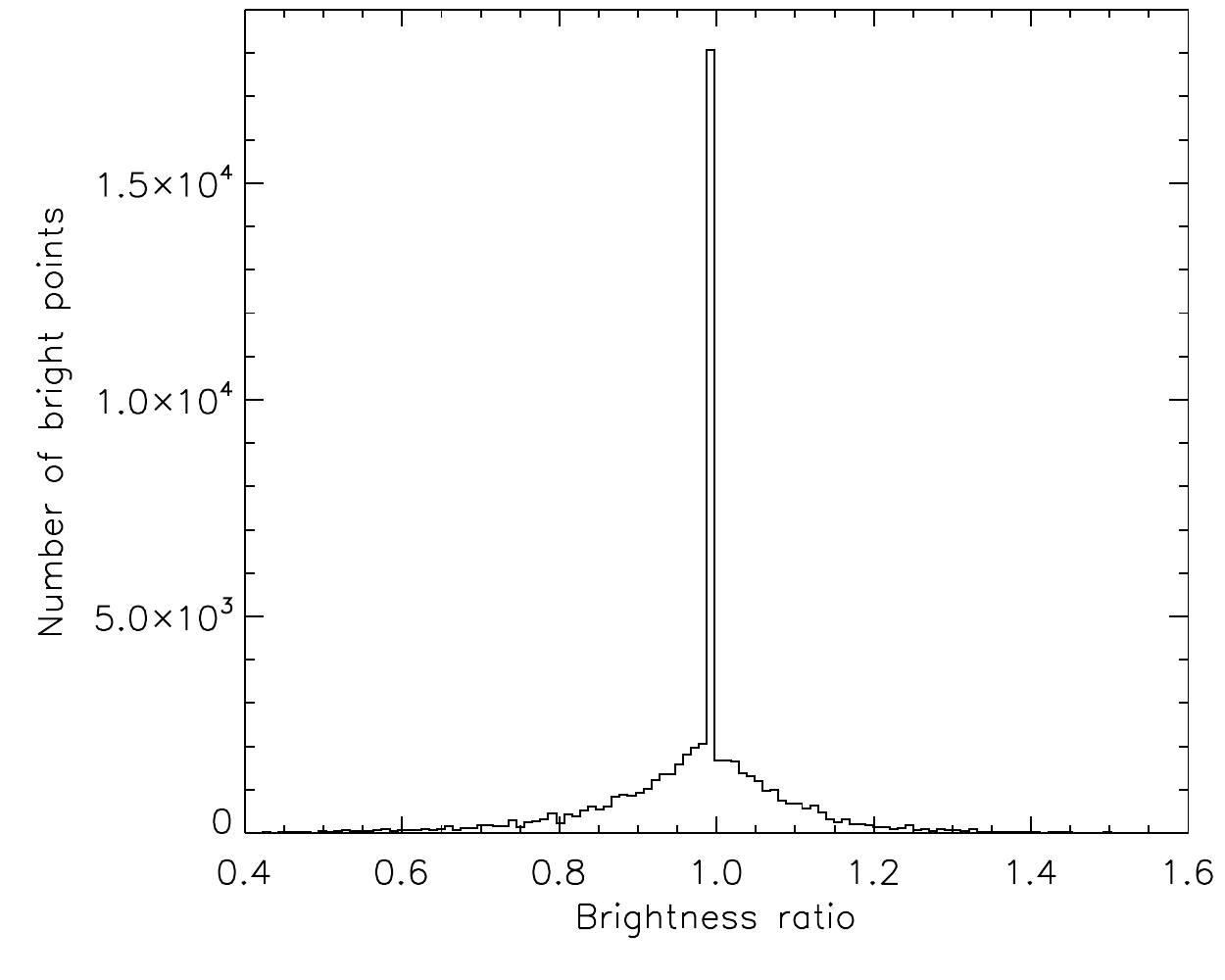}
\end{center}
\caption{Histogram of brightness ratio for bright points detected in $\sim$200 LASCO C2 images, and the same bright points detected in the same images, but rebinned to size $512\times512$. This is a test of the validity of comparing coincident bright points in \Btot\ and \pB\ images.}
\label{testimagerebin}
\end{figure}

One important source of systematic error is the influence of different pixel scales on the pixel identification routine and the resulting relative brightness between pixel groups. Most \pB\ images have twice the plate scale of the \Btot, that is, the \pB\ images are of size $512\times512$ whilst the \Btot\ images are $1024\times1024$. To allow for this, the procedures for isolating and identifying bright points in the \pB\ images use a sliding window median of 5 pixels squared (rather than the 11 pixels squared of the \Btot\ images). This correction to the procedure still has a potential of introducing a systematic error. This is tested by detecting points in many \Btot\ images, then rebinning the same images to size $512\times512$ and applying the \pB\ detection procedures. A histogram of the brightness ratio of detected points between the original and rebinned images is shown in figure \ref{testimagerebin}. The distribution is very strongly peaked at 1, with a median absolute deviation of 4\%, validating the choice of sliding window size.

\begin{figure}[]
\begin{center}
\includegraphics[width=7.0cm]{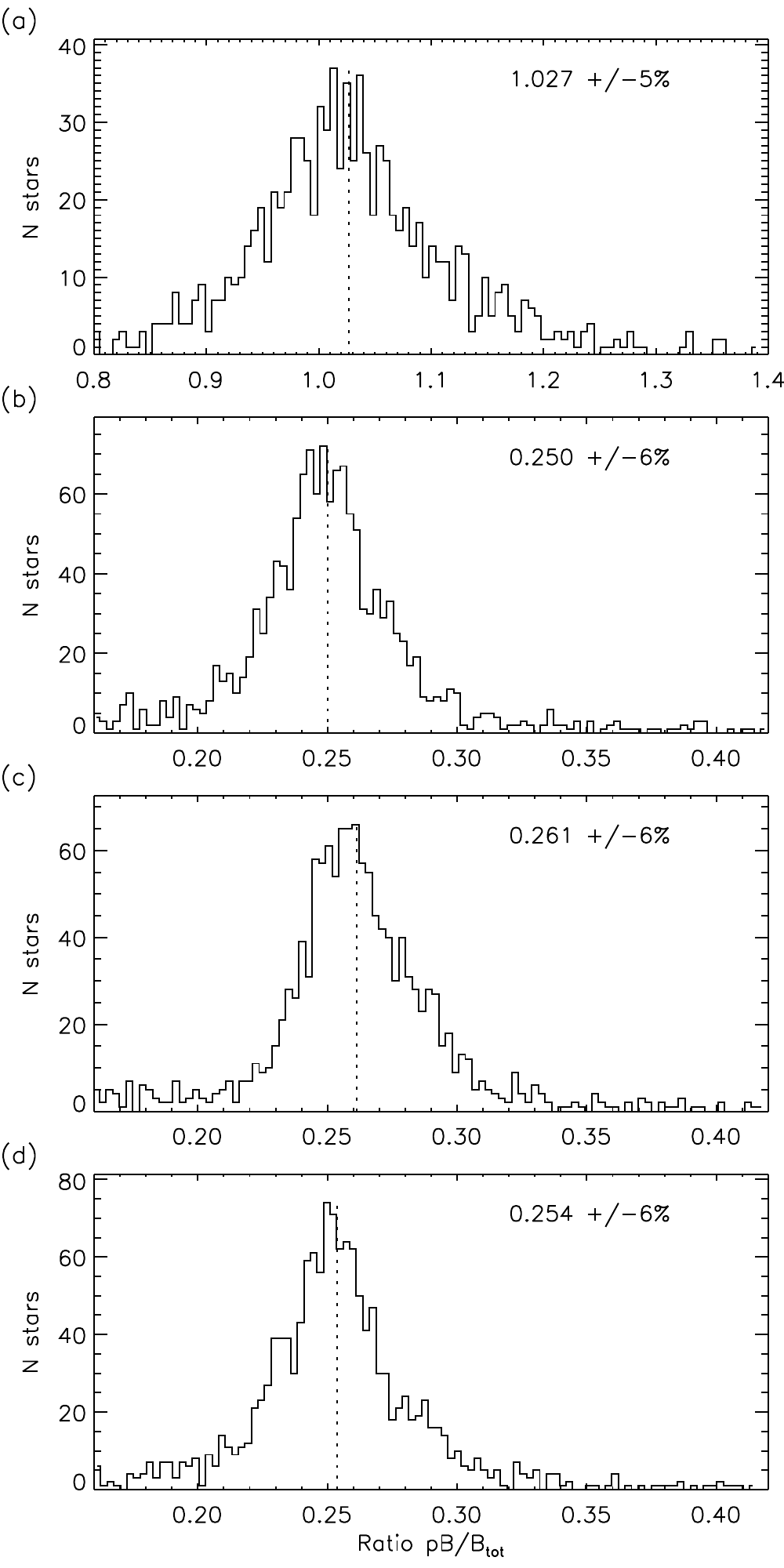}
\end{center}
\caption{Histograms of ratio of \pB\ star brightness to \Btot\ star brightness for polarizer states (a) clear, (b) $-60$\de, (c) 0\de, and (d) $+60$\de. The vertical dotted line shows the median ratio, and the annotation give the median and median absolute deviation of each distribution.}
\label{disppb}
\end{figure}

Figure \ref{disppb} show histograms of the ratio of the star's brightness in \pB\ compared to \Btot, over years 2000-2015. Note that high thresholds (which results in only the brightest stars being used for this part of the analysis), and the stringent criteria for associating bright points in \pB\ images to star tracks in \Btot\ images (distance of less than 1 pixel) has resulted in a relatively low number of stars. The median and median absolute deviations of ratios over all time for each polarizer state are listed in table \ref{tablepb}. The errors are around 5\%. These calibration factors differ only a little from the standard values used in the Solarsoft procedures, which are also listed in table \ref{tablepb}. The Solarsoft values are all within the margin of errors of the star values. Note also the small difference in calibration factor for the 0\de\ polarizer position compared to the other two angles, although this is still within the margin of errors. 

\begin{table}[h]
\begin{center}
\begin{tabular}{ c | c | c | c }
{\pB\ state} & {This work} & {Error (\%)} & {Current Solarsoft}\\
\hline
Clear & 1.027 & 5 &  1.0\\
$-60$\de & 0.250 & 6 & 0.25256\\
$0$\de & 0.261 & 6 & 0.25256\\
$+60$\de & 0.254 & 6 & 0.25256\\
\end{tabular} 
\caption{Correction factors for the LASCO C2 polarizer observations.}
\label{tablepb}
\end{center}
\end{table}

The low number of stars prevent the investigation of any variance of the correction factors across the image, that is, we are unable to test the vignetting of the \pB\ image. For the same reason, we are unable to make a detailed investigation of any variance in the \pB\ calibration factors over time, although values for the first few years after 2000 are almost identical to values taken in the last few years before 2015, suggesting no significant change. For the remainder of this work, we divide the LASCO C2 \pB\ polarizer images (the `clear' polarization observations are not used) by the correction factors listed in table \ref{tablepb}, then apply the \Btot\ calibration factors of \citet{colaninno2015}.

\begin{figure}[]
\begin{center}
\includegraphics[width=10.0cm]{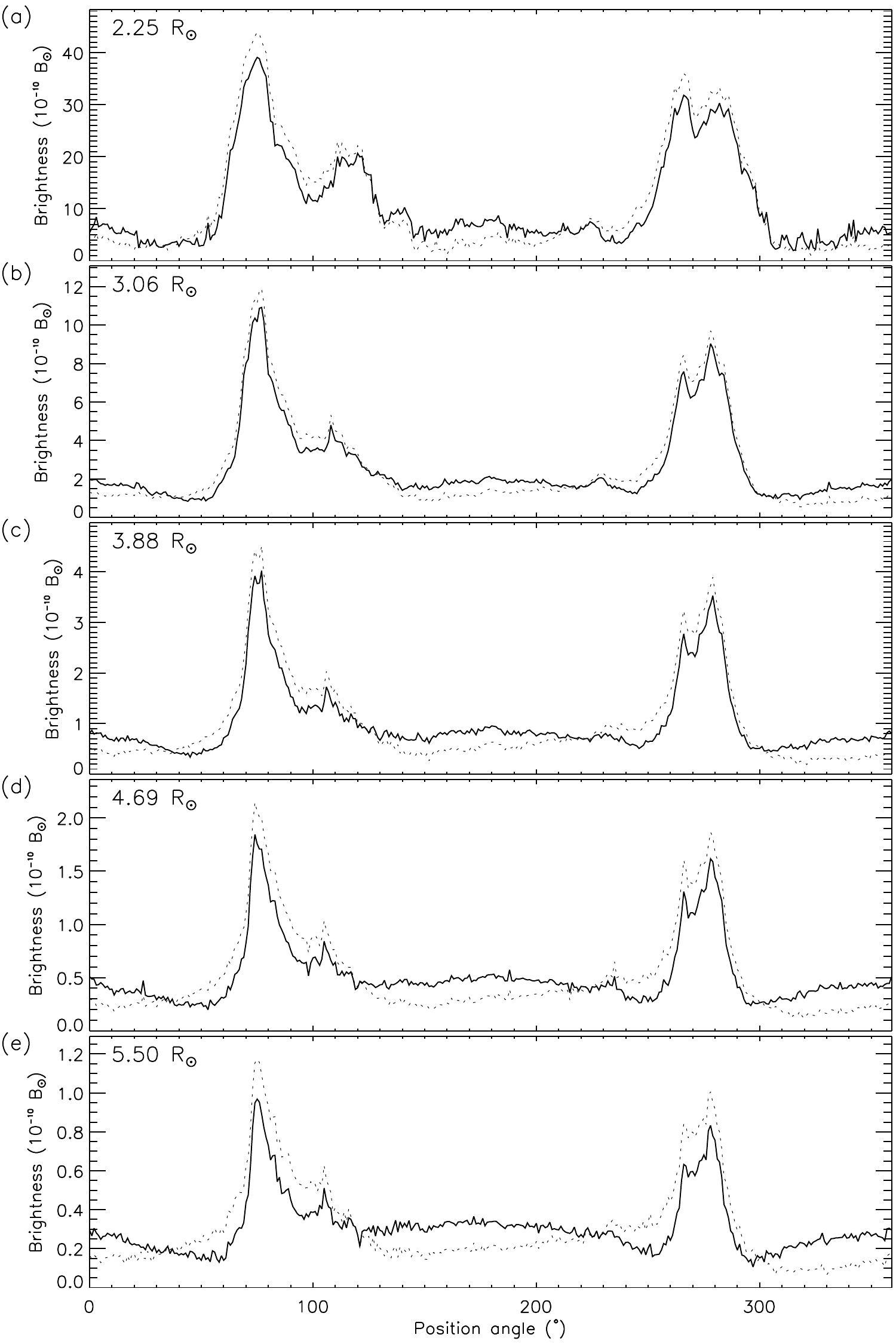}
\end{center}
\caption{Comparison of LASCO C2 \pB\ values calibrated using the old Solarsoft factors (solid line) with the new factors estimated from stars listed in table \ref{tablepb} (dotted line). Cuts at constant heights are shown as function of position angle.}
\label{newpb}
\end{figure}

Although the new factors appear very close to the old Solarsoft factors, there is considerable difference in the resulting \pB\ images combined from the sequences. Figure \ref{newpb} show profiles of \pB\ at various heights as a function of position angle, calculated using the old and new factors. Two suspect features of \pB\ images using the old factors are the unexpected increase of \pB\ over the poles, and strange dips near the equatorial streamers. This is most apparent during solar minimum and at larger heights. These features are eliminated using the new correction factors, leading to broad regions of minimum brightness over the poles and no dips next to the streamers. A paper of relevance to this part of the work is \citet{moran2006}, who derived new Mueller matrix formulations to correct for a phase error in LASCO C2 and C3 \pB\ observations. We have not attempted to revisit their analysis using the new C2 calibration factors, but may attempt this in a future work.

\subsection{Approximation of $B_k$ using observed $pB$}

\begin{figure}[]
\begin{center}
\includegraphics[width=8.0cm]{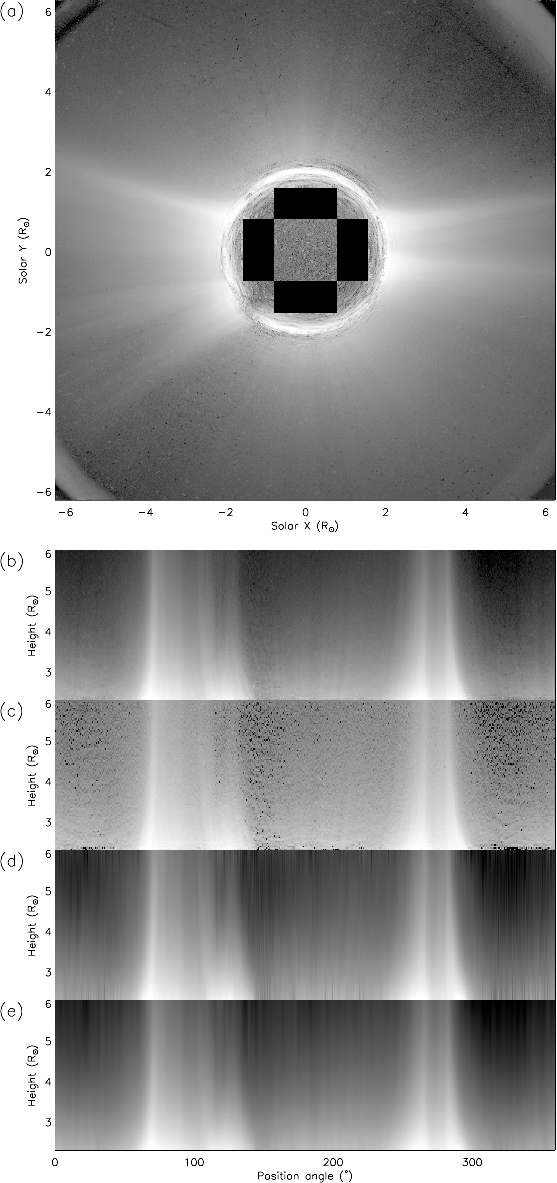}
\end{center}
\caption{Stages in the conversion of LASCO C2 \pB\ image to \Bk. (a) \pB\ observation of 2007/03/21 (log brightness). (b) The same image transformed to polar coordinates. (c) Electron density in the plane of sky gained from inversion of the \pB\ image using a local spherical symmetry (see text). (d) Electron density fitted to a polynomial function of height. (e) \Bk\ calculated by integrating the electron density along appropriate lines of sight, again using local spherical symmetry.}
\label{lascopb}
\end{figure}

\begin{figure*}[]
\begin{center}
\includegraphics[width=12.0cm]{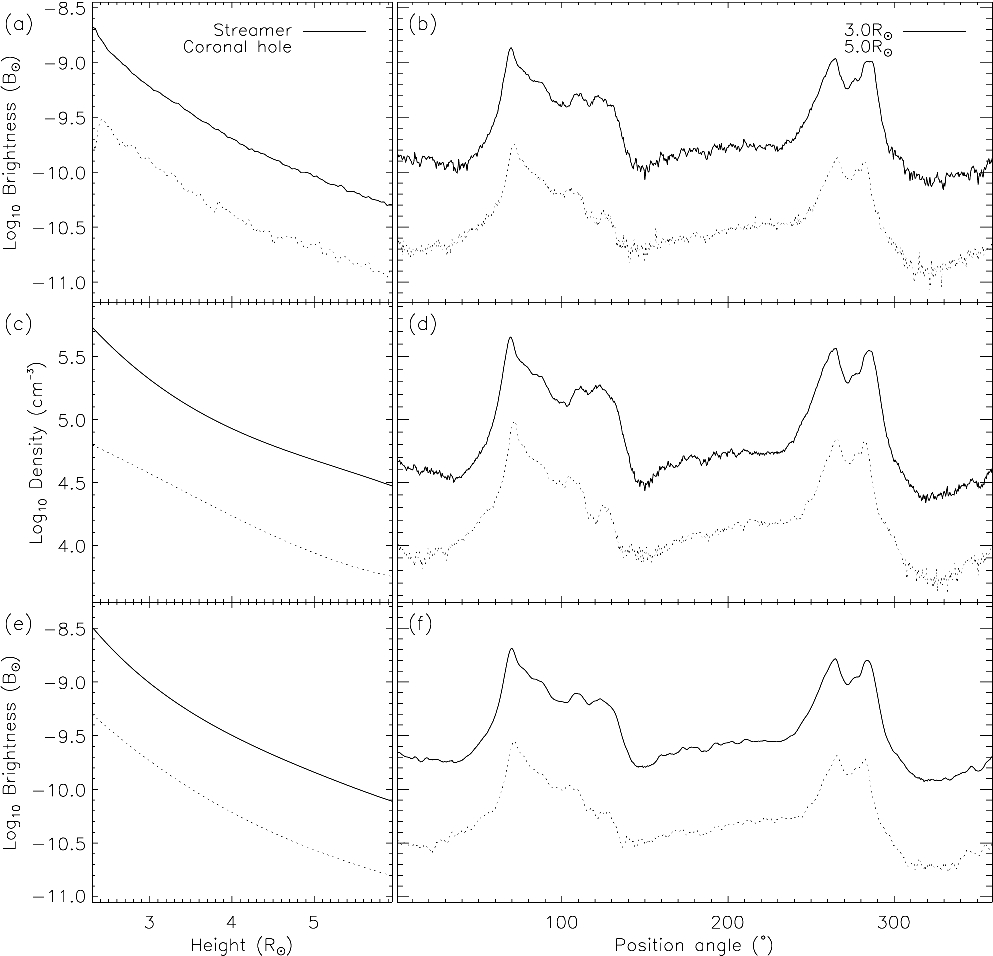}
\end{center}
\caption{Line plots illustrating the conversion of \pB\ to \Bk\ for selected `slices' of the data. The left column shows (a) observed \pB, (c) electron density and (e) \Bk\ profiles vs. height for a streamer (solid line) and coronal hole (dotted). The right column shows (b) observed \pB, (d) electron density and (f) \Bk\ vs. position angle at heights of 3.0\Rs\ (solid line) and 5.0\Rs\ (dotted line).}
\label{lascopb2}
\end{figure*}

LASCO C2 \pB\ sequence files are opened and processed using the standard LASCO Solarsoft procedures, with the new calibration factors listed in table \ref{tablepb}. An example \pB\ observation is shown in figure \ref{lascopb}a. This image is transformed into polar coordinates, as shown in figure \ref{lascopb}b. Figures \ref{lascopb2}a,b show selected radial and angular profiles of \pB. For each position angle, the radial profile of \pB\ is assumed to arise from a locally spherically symmetric distribution of electron density, enabling an inversion of \pB\ into density, as shown in figure \ref{lascopb}c. The procedure used is very similar to that described by \citet{quemerais2002}. This procedure is sensitive to errors in the observations and can give a rather unsmooth distribution of density (i.e. noise is amplified). This is alleviated by fitting each radial profile of density to a second-degree polynomial in $\log _{10}$ space. Fitting the radial profiles across all position angles gives three fitting parameters as a function of angle. These are smoothed across a few degrees of position angle and the smoothed parameters used to reconstruct the radial density profiles. The resulting fitted density is shown in figure \ref{lascopb}d and figures \ref{lascopb2}c,d. Fitting the density with this procedure helps to reduce errors due to noise greatly, and also reduce the effect of CMEs in the calculated density. Finally, the density is integrated along appropriate lines of sight (given the geometrical factors given by e.g. \citet{quemerais2002}) resulting in an estimate of \Bk, as shown in figure \ref{lascopb}e and figures \ref{lascopb2}e,f. The intrinsic errors in assuming a spherically-symmetric distribution of density to calculate \Bk\ is discussed in a following section.

\subsection{Calculation of calibration factors}

\begin{figure}[]
\begin{center}
\includegraphics[width=8.5cm]{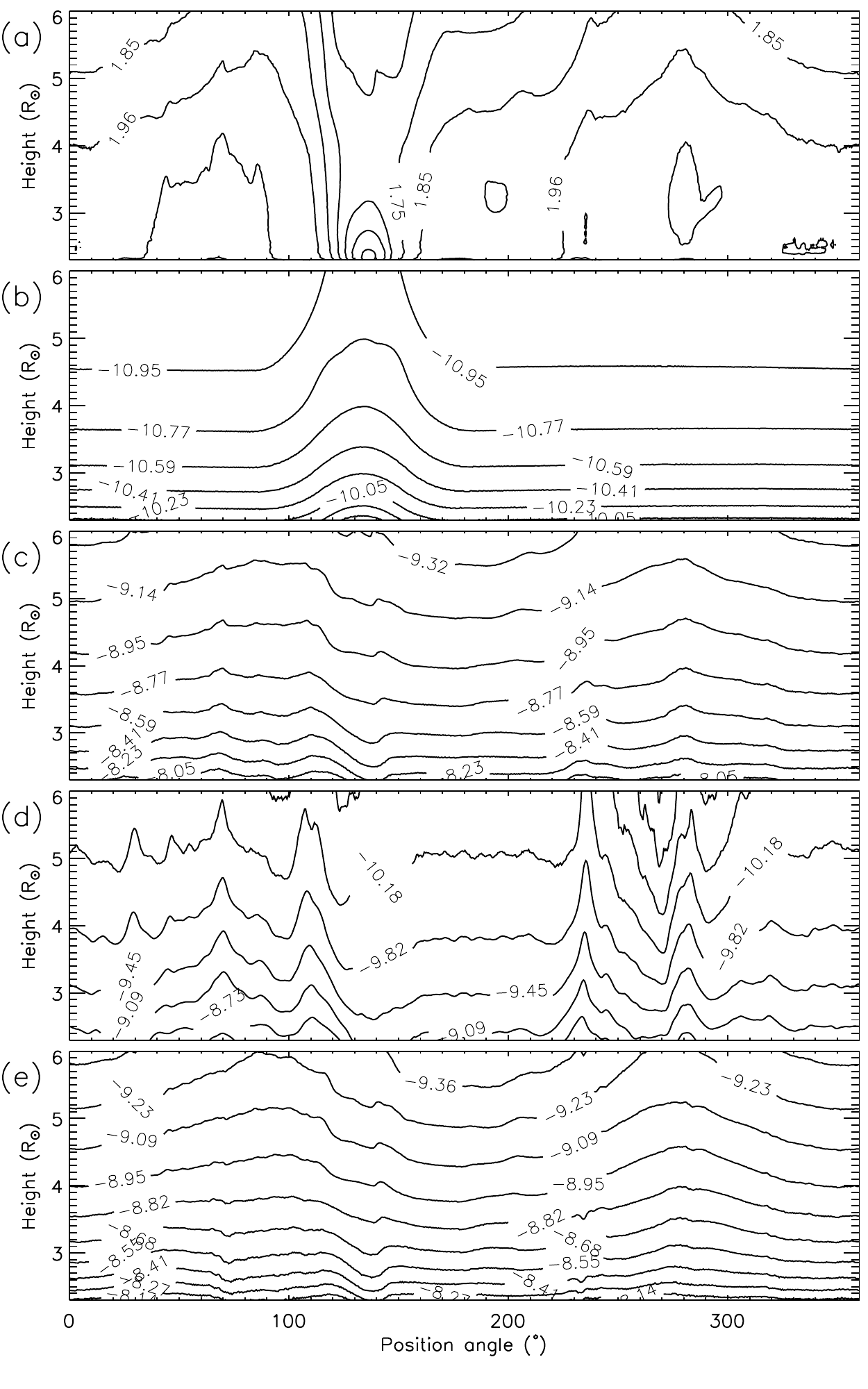}
\end{center}
\caption{Series of plots illustrating the calculation of individual calibrated backgrounds. (Note that these calibration backgrounds are combined over a long time period ($\sim10$ days) for final use). (a) The median image calculated from 8 hours of uncalibrated LASCO C2 total brightness observations during 2007/03/21, all normalized by exposure time and DST-processed (quiescent component). (b) Calibrating and vignetting image as given by the standard LASCO Solarsoft routines. This is a multiplicative factor for calibration of total brightness images. (c) Result of applying the calibration image to the median image (product of (a) and (b)). (d) A \Bk\ image calculated from a \pB\ observation made during the 8 hour window. (e) Subtraction of \Bk\ from the calibrated median image (c). This is a calibrated background for subtraction from total brightness images i.e. multiplying the total brightness image of (a) by (b) and subtraction of the background (e) will result in an image identical to the \Bk, (d). }
\label{calfiles}
\end{figure}

The procedure has been applied to all LASCO \pB\ observations, resulting in a large set of \Bk\ images in polar coordinates. For the time of each \pB\ observation, the quiescent component total brightness files within $\pm4$ hours are identified and opened. From these, a time-median image is formed (i.e. for each spatial pixel a median over time is calculated). Such an image is shown in figure \ref{calfiles}a. This image is vignetted and radiometrically calibrated to mean solar brightness (\msb) units using the standard LASCO solarsoft procedures and the radiometric calibration factors of \citet{colaninno2015}. Figure \ref{calfiles}b shows a vignetting calibrating image. The raised profile centered on position angle $\sim$135\de\ is to correct for the effect of the occulter's pylon in the LASCO C2 field of view. Multiplying the uncalibrated total brightness image by this vignetting calibrating image gives the calibrated image shown in figure \ref{calfiles}c. Finally, subtraction of the \Bk\ image gained from the \pB\ observation (shown in figure \ref{calfiles}d), gives the image of figure \ref{calfiles}e. This is an individual calibrated background, calculated using a single \Bk\ image and $\sim$8 hours worth of DST-processed total brightness images. The next step is to combine many such images over a long time period, and a critical test is the stability of the background over long time periods.

\begin{figure}[]
\begin{center}
\includegraphics[width=8.5cm]{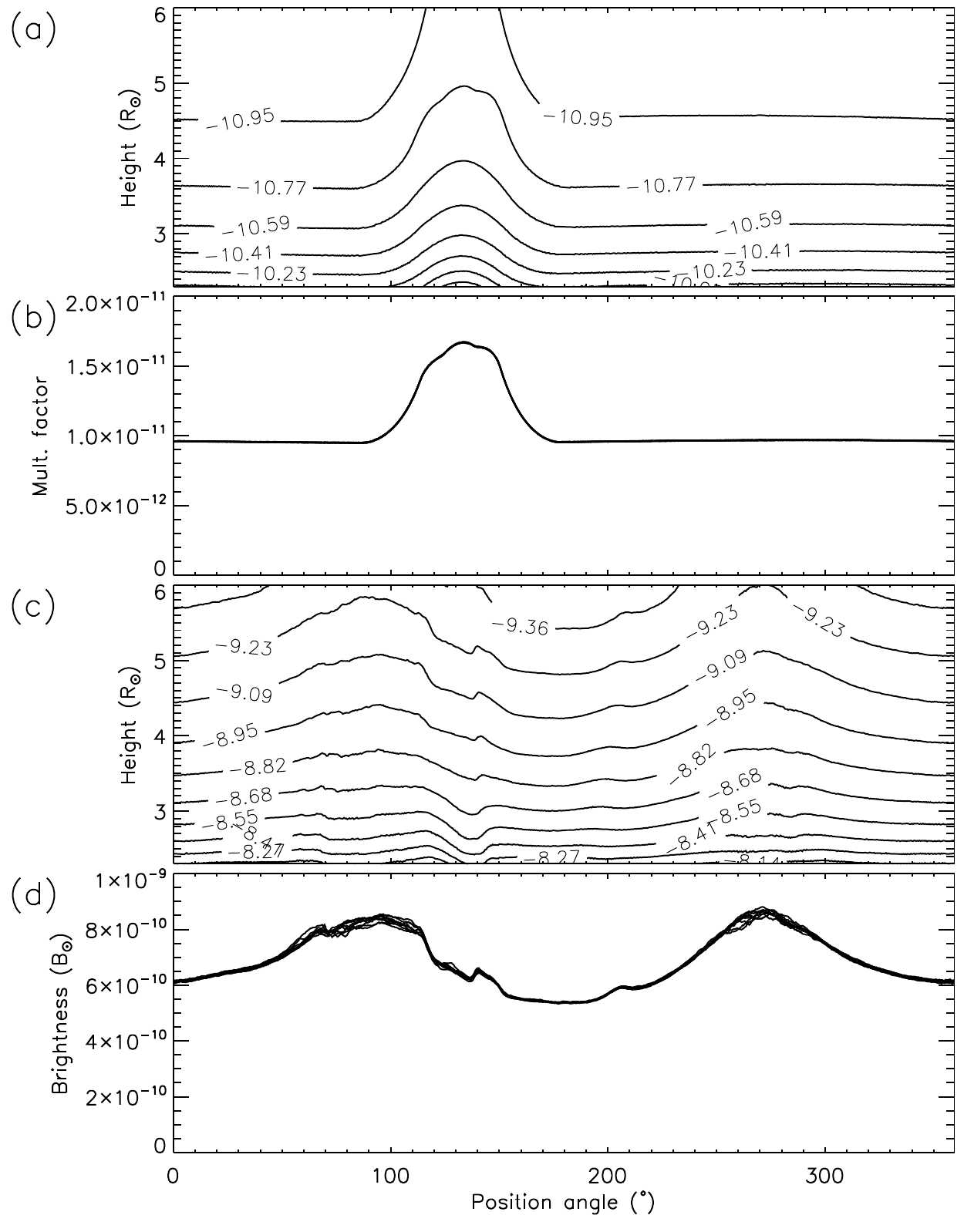}
\end{center}
\caption{(a) Calibrating/vignetting image, median over the 10 days during 2007/03. (b) Individual calibrating/vignetting profiles vs. position angle calculated for all \pB\ observations over the 10 days, plotted for a height of 5.0\Rs. The individual functions are almost identical, giving the impression of a single thick line. (c) Calibrated background image, median over the 10 days. (d) Individual calibrated background profiles vs. position angle, at a height of 5.0\Rs, over the course of 10 days. In this case, there is a little more variation in the values.}
\label{calfileslt}
\end{figure}

\begin{figure}[]
\begin{center}
\includegraphics[width=8.0cm]{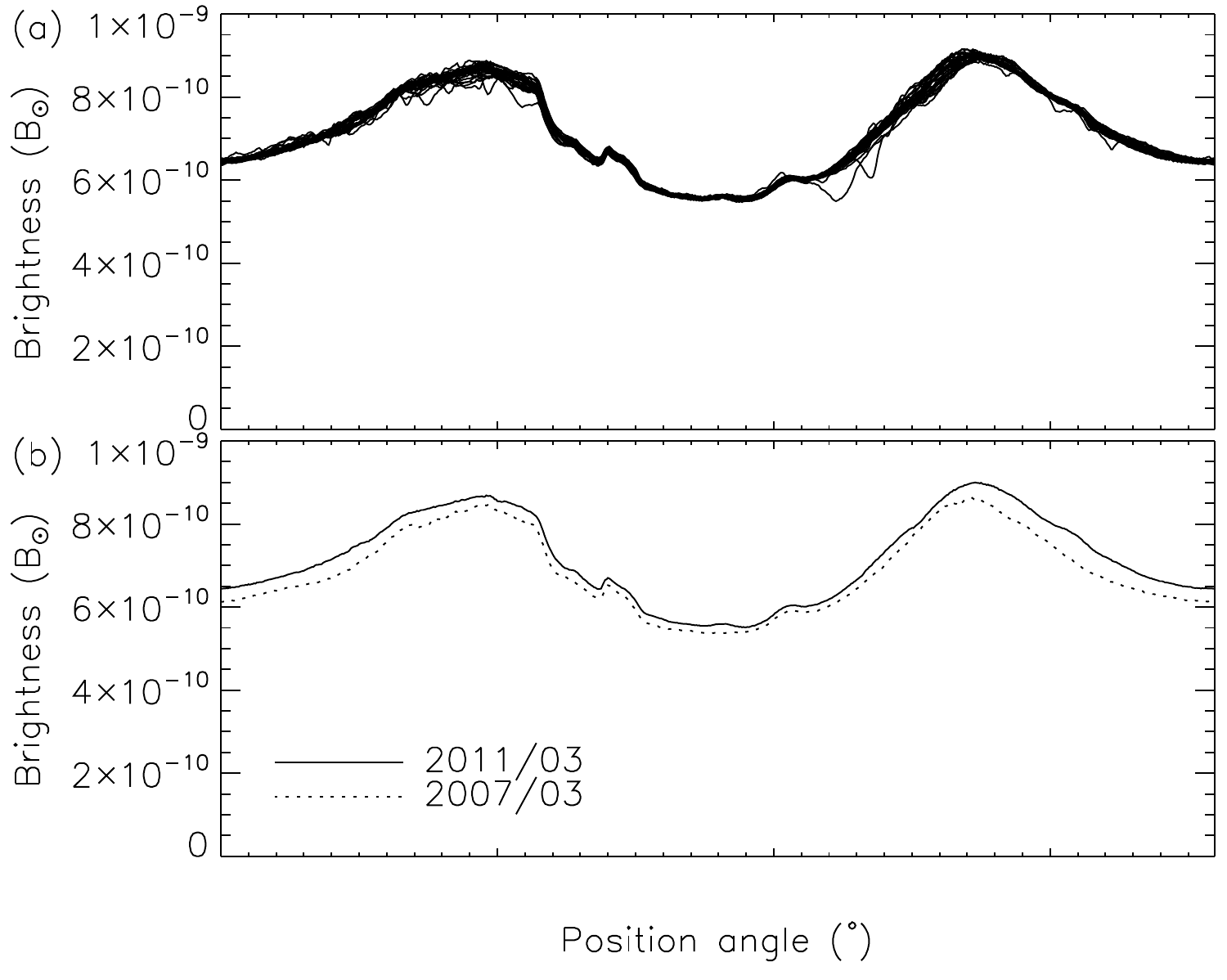}
\end{center}
\caption{(a) Individual calibrated background profiles vs. position angle, at a height of 5.0\Rs, over the course of 10 days during 2011/03. Note the increased variance compared to 2007/03. (b) Comparison of the median calibrated background profiles vs. position angle for 2011/03 (solid line) and 2007/03 (dotted). }
\label{calfileslt2}
\end{figure}

To create a suitable final calibration image, many calibrating and background subtraction images (as illustrated in figure \ref{calfiles}) are combined over $\pm$5 days from the date of interest. Figure \ref{calfileslt} shows the resulting images and illustrates how stable the calibration and background images are over the course of $\sim$10 days, centered on 2007/03/21. The vignetting calibrating function (figures \ref{calfileslt}a,b) changes very little over the period, as expected. The calibrated backgrounds shown in figures \ref{calfileslt}c,d also remain remarkably stable. This is in part due to the use of the DST on the total brightness images and the use of an 8-hour median of total brightness quiescent component images. But in this case it is also due to the low solar activity during 2007. CMEs can cause large differences between the \Bk\ and DST-processed quiescent images, resulting in larger errors in the individual calibrated backgrounds. Fortunately, this is easily remedied by taking the median over the 10 day period. Figure \ref{calfileslt2}a shows many individual background profiles vs. position angle at a height of 5.0\Rs\ for a 14 day period over 2011/03. The variance is again quite small for most profiles. One of the profiles varies considerably but does not skew the final median background, shown in figure \ref{calfileslt2}b next to the 2007/03 background. The backgrounds, separated by 4 years and from two extremes of the solar cycle are very similar in profile. The 2011 profile is consistently larger then the 2007 profile by a few percent.

\begin{figure}[]
\begin{center}
\includegraphics[width=8.0cm]{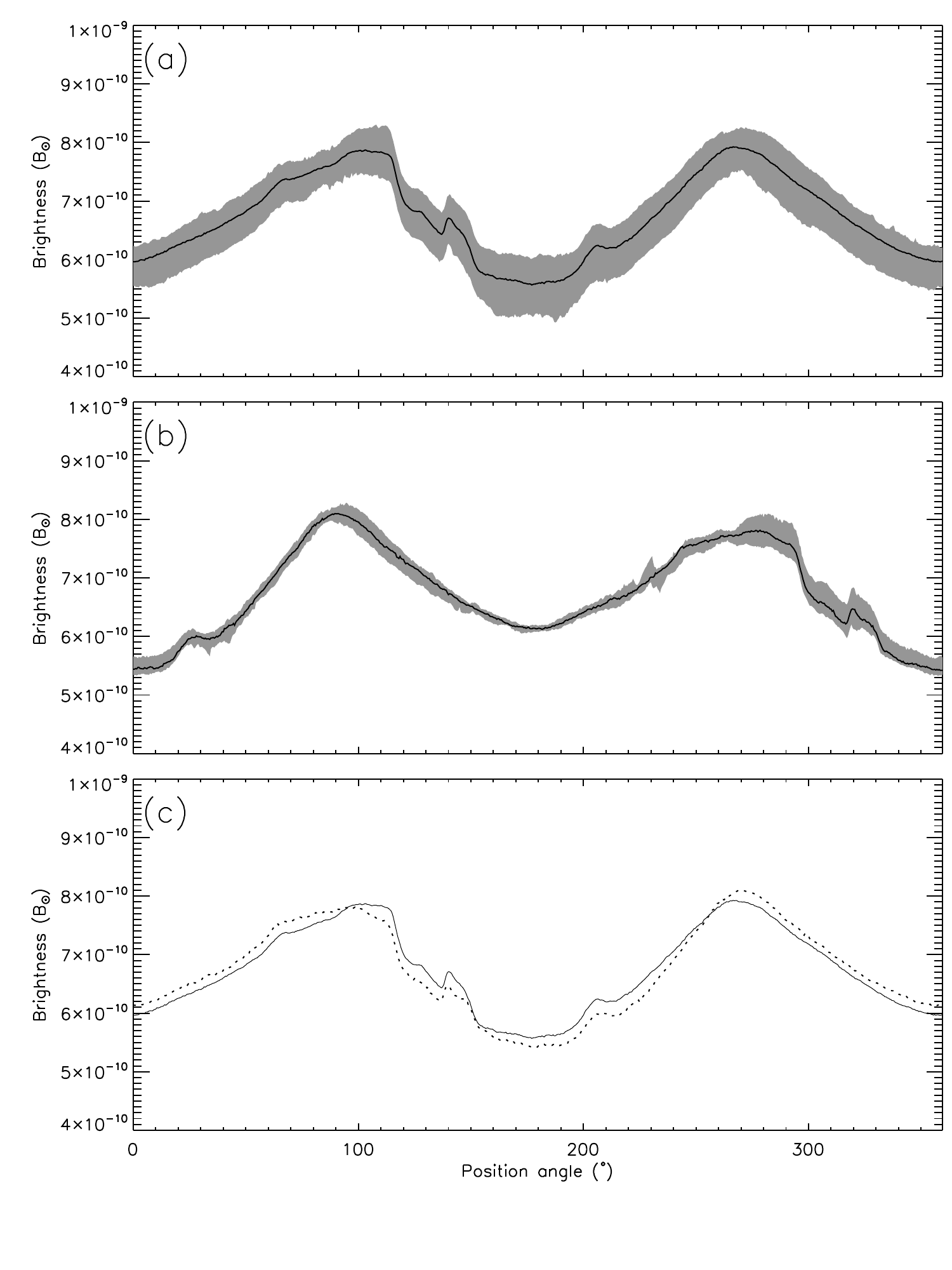}
\end{center}
\caption{Background profiles vs. position angle at a height of 5.0\Rs. (a) Background profile calculated for the long period 1999/12/12 to 2003/07/09. The shaded area shows the minimum ($1^{st}$ percentile) and maximum ($99^{th}$ percentile) values over this time, the solid line shows the median. SOHO did not make any roll maneuvers during this time, maintaining the LASCO C2 image vertical at angles close to solar north. (b) Background profile for period 2003/07/11 to 2003/10/07, when SOHO was in a $\sim$180\de\ roll position. (c) Comparison of the medians over both periods with the 2003/07/11-2003/10/07 profile (dotted line) shifted by 180\de.}
\label{testlt}
\end{figure}

The stability of the subtracting background images is important as a test of their validity. The background should contain the brightness of the F-corona (\Bf) plus any instrumental stray light not removed by the DST. Figure \ref{testlt} compares the background profiles vs. position angle at a height of 5.0\Rs\ for two different periods. The SOHO roll angle differs by $\sim$180\de\ between these two periods. From 1999/12/12 - 2003/07/09 the SOHO roll angle aligned the LASCO C2 image vertical to within a few degrees of solar north. The median and minimum/maximum values of the background (at the 1$^{st}$ and 99$^{th}$ percentiles) for this long period are shown in figure \ref{testlt}a. Figure \ref{testlt}b shows the same for the period 2003/07/11 to 2003/10/07, when the roll angle was at $\sim$180\de\ compared to the previous period. The profiles are obviously different due to the different roll angles, thus confirming the presence of instrumental effects on this background. The \Bf\ component should remain unchanged with rotation, and is likely smooth. Figure \ref{testlt}c compares both medians on the same plot, with the 2003/07/11-2003/10/07 profile shifted by 180\de. The profiles are very similar. The smaller-scale structures and the small differences between the two broad peaks at 90 and 270\de\ are likely instrumental effects because they rotate with the spacecraft rolls. It is important therefore to use the correct long-term background for the correct roll period. At times when SOHO makes roll manouevers, the 10-day `sliding window' for creating median long-term backgrounds is truncated, and a new sliding window is begun from the start of the new roll position. 

\begin{figure}[]
\begin{center}
\includegraphics[width=8.0cm]{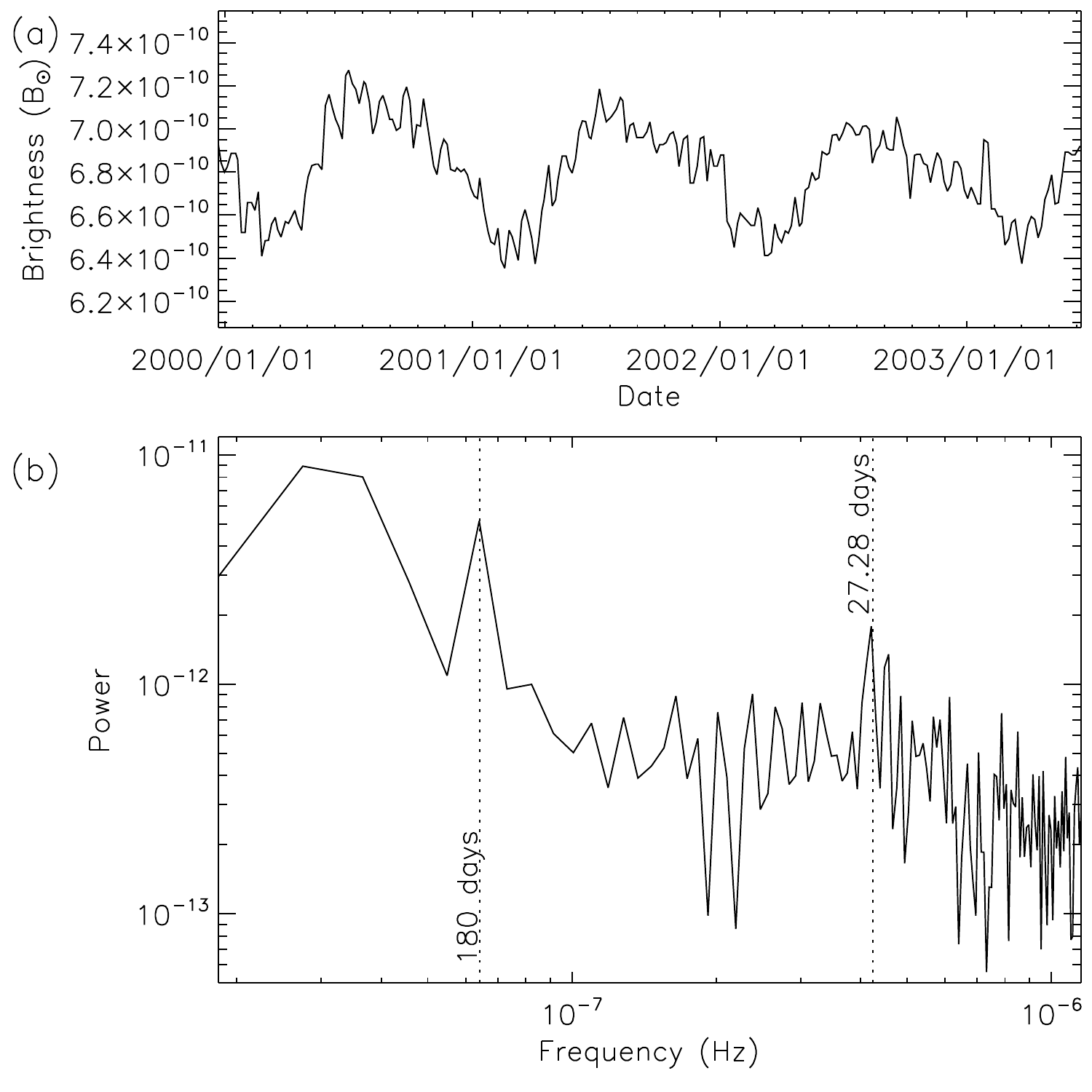}
\end{center}
\caption{(a) Variation in the calibrated background over a long time period. The value is shown for a single point at position angle 125\de\ and height 5.0\Rs. (b) Fourier power spectrum of (a). Peaks are found at 180 days (SOHO orbital period) and at 27 days (Carrington rotation period).}
\label{calosc}
\end{figure}

As a last point for this section, figure \ref{calosc}a shows the variation of one point in the background subtraction images over the 1999/12/12-2003/07/09 period (a period with no roll manouevers). A close to yearly oscillation is seen, as well as shorter-term oscillations which seem periodic. Figure \ref{calosc}b shows the Fourier power spectrum of this time series. Peaks are seen at 360 and 180 day periods, probably due to the SOHO orbit. There is another distinct peak near the 27 day period corresponding to the Carrington rotation. These oscillations, particularly the variation due to the SOHO orbit, accounts for most of the variation in the long-term backgrounds.

\subsection{Application of calibration to LASCO C2}

Figure shows the application of calibration and background subtraction to the observation of 2007/03/21 11:26 (shown without DST processing and calibration in figure \ref{figsep}a). Following the procedures of this section, this image approximates \Bk. For the four points indicated by diamonds the brightness values over a 14 day period is plotted in figure \ref{wlcal}. Overplotted with triangles are the \Bk\ values calculated by inversion from the daily \pB\ observations. In general, we should not expect the agreement between the two to be perfect. The \pB\ observations may contain CMEs which have not been removed using DST due to the low cadence. Additionally, the background subtraction images are calculated over a 10-day sliding window. Perhaps the biggest source of error is in the approximation of \Bk\ from the observed \pB, discussed in the following section. The large increase and decrease in brightness over the course of 4 days in figure \ref{wlcal}c is due to the movement of a streamer into and out of the line of sight.

\subsection{Error}

In the absence of cross-calibration with other coronagraphs there is a LASCO radiometric calibration error of $\sim$9\%\ to the brightness values, determined by observations of stars by \citet{colaninno2015}. This is an overall calibration uncertainty, not to be confused with pixel-to-pixel noise. Problems are also found in isolated images which are usually caused by a shower of energetic particles hitting the detector. Such noisy images can cause big problems for the DST and an attempt is made to identify and discard such images during batch processing. The point filter helps with images which contain only limited degradation.

The biggest weakness of the method is the approximation of \Bk\ from \pB\ using a local spherically symmetric distribution of density. The worst case would be for regions of the corona where both streamers and low-density coronal holes lie along the line of sight. In this case, the distribution of density is far from spherically-symmetric and our inversion will lead to errors. The most accurate inversion will occur in the large polar coronal holes and our estimation of \Bk\ will be best in these regions. Errors in the calculation of \Bk\ lead to errors in the formation of the individual background images (i.e. an incorrect value of \Bk\ is subtracted from total brightness images leading to errors in the resulting backgrounds). This is most obvious within or near streamers, and can be seen as narrow radial deviations from the smoother long-term median backgrounds. Luckily, such localised deviations are avoided by the creation of long-term medians. Thus, if the \Bk\ inversion errors due to using spherical symmetry are likely short-lived errors, they will be removed from the final background images. An estimate of the amount of error introduced by the procedure is given by the amplitiude of the 27 day periodicity seen in figure \ref{calosc}. For this example, at a height of 5.0\Rs\ and position angle 125\de, the error is around $2 \times 10^{-12}$\msb, or $\sim$2\%\ of the mean background level. 

\section{Cross calibration of LASCO C2 and SECCHI COR2 A}
\label{c2cora}

During 2007/03, SOHO and STEREO A \& B were very close in position, therefore LASCO C2 and the SECCHI CORs were viewing the corona from very similar angles. This gives an opportunity to compare brightness values and to cross-calibrate. The corona is close to a solar minimum configuration with the main streamer belts near the equator in the west and east, and the low activity also aids in cross-calibration. In this section, correction factors will be found to apply to the individual SECCHI COR2 A \pB\ sequence observations in order for them, after combination, to more closely match the LASCO C2 \pB.

\begin{figure}[]
\begin{center}
\includegraphics[width=8.0cm]{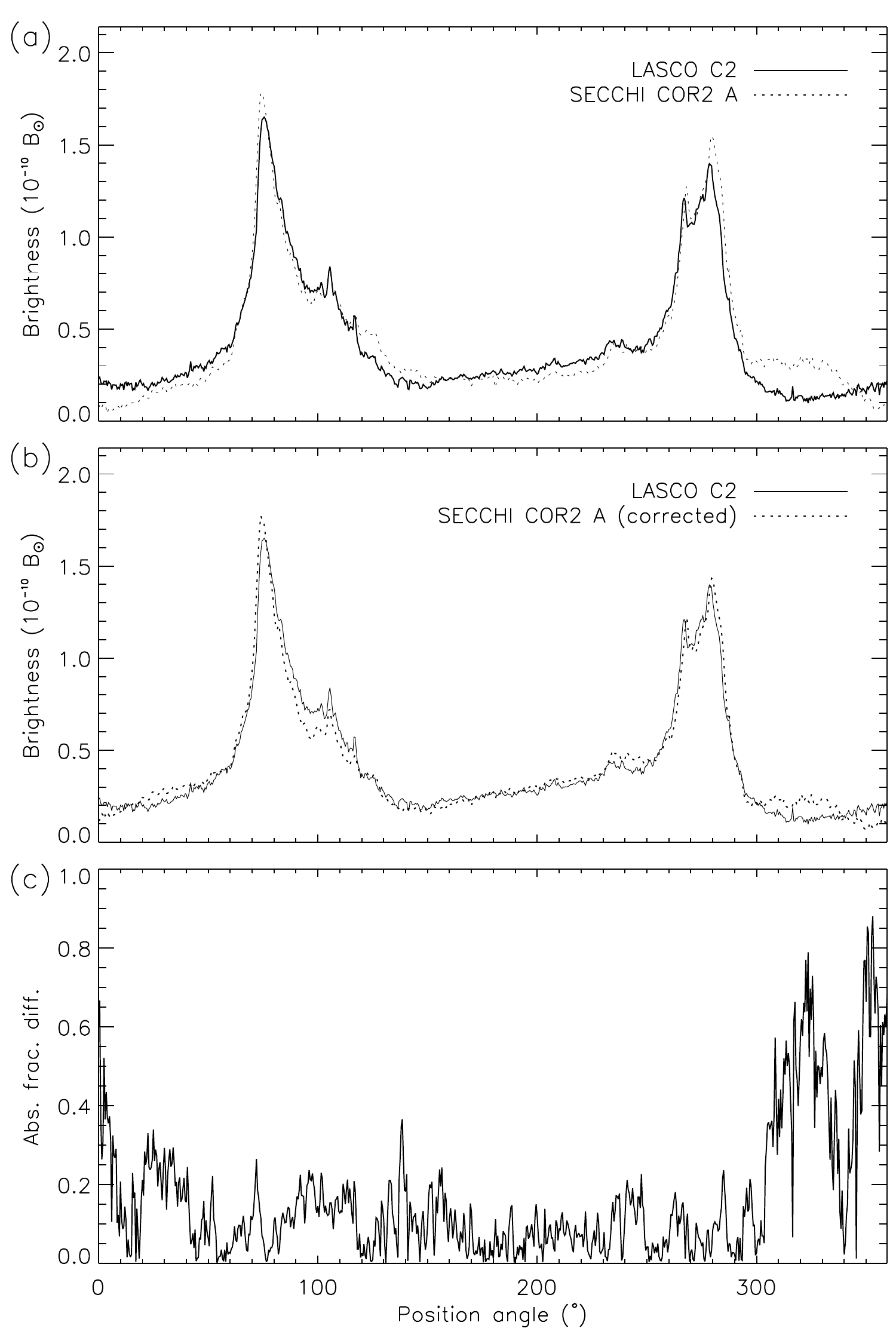}
\end{center}
\caption{(a) Comparison of LASCO C2 (solid line) and SECCHI COR2 A (dotted) polarized brightness at a height of 5\Rs\ as a function of position angle for observations within a few minutes of 2007/03/20 21:00. (b) As (a), but with correction factors applied to COR2 A. (c) Absolute fractional difference of the two profiles of (b) as a function of position angle}
\label{crosscal2}
\end{figure}

Figure \ref{crosscal2}a shows \pB\ profiles vs. position angle at a height of 5\Rs\ for both LASCO C2 and SECCHI COR2 A for observations made close to 2007/03/20 21:00. The differences between the profiles makes any direct analysis combining both instruments unreliable. This disagreement is far worse without the corrections to the LASCO C2 \pB\ correction factors found above using stars. Over the period of 2007/03/16-2007/03/29, all LASCO C2 \pB\ sequences are calibrated and recorded. For every LASCO C2 \pB\ sequence, the sequence closest in time made by COR2A is identified. The observations made at the 3 polarizer angles of 0, 120 and 240\de\ are opened using the Solarsoft Secchi\_prep.pro software, allowing rotation and radiometric calibration, but the three observations are not combined. Correction factors are applied independently to all three polarizer sequences before applying the inverse Mueller matrix and combining, with the aim of minimizing the difference between \pB\ in C2 and COR2A. The comparison set of data is limited to heights between 4 and 6\Rs. The best fit is found when the three polarizer angles of COR2A, 0, 120 and 240\de\, are multiplied by correction factors of 0.960, 0.977, and 0.972 respectively. Figure \ref{crosscal2}b shows a corrected COR2A profile in comparison to LASCO C2, as a function of position angle.

The corrected COR2 A brightness profile results in a much closer match with the LASCO C2 profile, as shown in figure \ref{crosscal2}b. Figure \ref{crosscal2}c shows the absolute fractional difference between the two instruments, post-correction. The decision was made to adapt the COR2 A brightness to match the LASCO C2 because the C2 instrument has been subject to far more detailed radiometric calibration in this work and others, particularly in-flight calibration using stars. For this reason, the LASCO C2 instrument is considered to be more accurate and used as a standard against which the COR2 A is calibrated. The crosscalibration factors are valid for heights between 4.0 and 6.0\Rs. Although COR2 A does observe to lower heights, a large stray light feature near the occulter in the center bottom of the images hinders an accurate analysis. At 4.0\Rs\ and above, this feature is avoided. For the purpose of the tomography method, using DST quiescent images, a single height at 5.0\Rs\ is sufficient. For a comparison of CME mass, however, it is desirable to maximize the overlapping height range.

\begin{figure}[]
\begin{center}
\includegraphics[width=8.0cm]{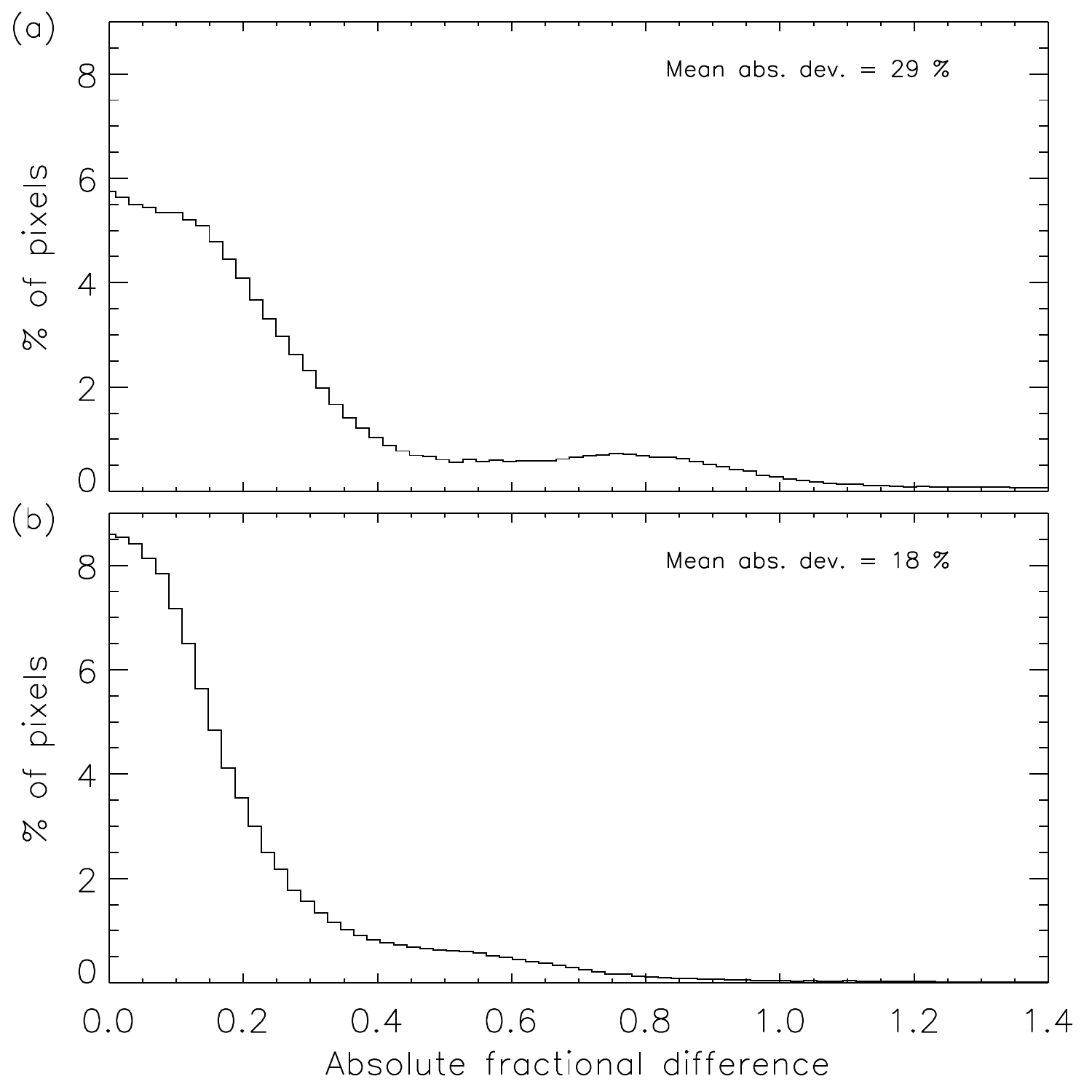}
\end{center}
\caption{Histograms of absolute fractional differences between LASCO C2 and SECCHI COR2 A \pB\ values for all position angles and heights for the 2007/03/16-29 period used for crosscalibration. (a) Without the correction procedure for COR2 A and (b) with the correction.}
\label{crosscal3}
\end{figure}

Figure \ref{crosscal3} shows the distribution of absolute fractional differences between LASCO C2 and COR2 A \pB\ without and with the correction for the whole 2007/03/16-29 period. The absolute fractional difference $d$ between two values $a$ and $b$ is
\begin{equation}
d=2\frac{|a-b|}{a+b}.
\end{equation}
With the correction the absolute fractional difference has a mean of 18\%, compared to 29\%\ without. The slightly different viewpoints of the two instruments may contribute to this difference. Across the many years of data the corrections will be applied to COR2 A using the values found for 2007/03. Unfortunately there is no other period where such a crosscalibration may be found without large uncertainties due to the large separation between spacecraft. At large separations, the best test for the calibration is made by tomographical reconstructions of brightness to give a distribution of emission or electron density. This comparison will be made in a following paper. 

\section{Cross calibration of SECCHI COR2 A and B}
\label{cor2ab}

At the heights of interest to this study (around 5\Rs), crosscalibration between the two SECCHI COR2 coronagraphs is made very difficult by the dominance of stray light features in the individual polarized brightness images of COR2 B. The individual images made at polarizer angles 0, 120 and 240\de\ compare badly with those of COR2 A. There is no simple cross-calibration solution as shown for LASCO C2 and COR2 A in the previous section. Attemps have been made involving analysing the differences between COR2 A and B over two weeks of observations at individual polarizer angles, calculating the mean of these differences over time to create correction images, and searching for a best fit between COR2 A and B by adjusting calibration factors for B's individual polarizer angle images. Several similar approaches based on the assumption that the stray light features may be subtracted or divided out from the individual polarizer images have been in vain. The best solution found is to calculate \pB\ images, and then calculate a long-term mean difference between COR2 A and B. This correction is then subtracted from the COR2 B \pB\ images. Additionally, corrective multiplicative factors have been calculated for the images at individual polarizer positions, used with the subtraction images. This approach, described below, is successful for the time period under study (2007/03/16-29), but is clumsy and unsatisfactory. There is no certainty that it is valid outside of the time of study, although a comparison of 3D emission using tomography will aid with this in \paperii.

\begin{figure}[]
\begin{center}
\includegraphics[width=8.0cm]{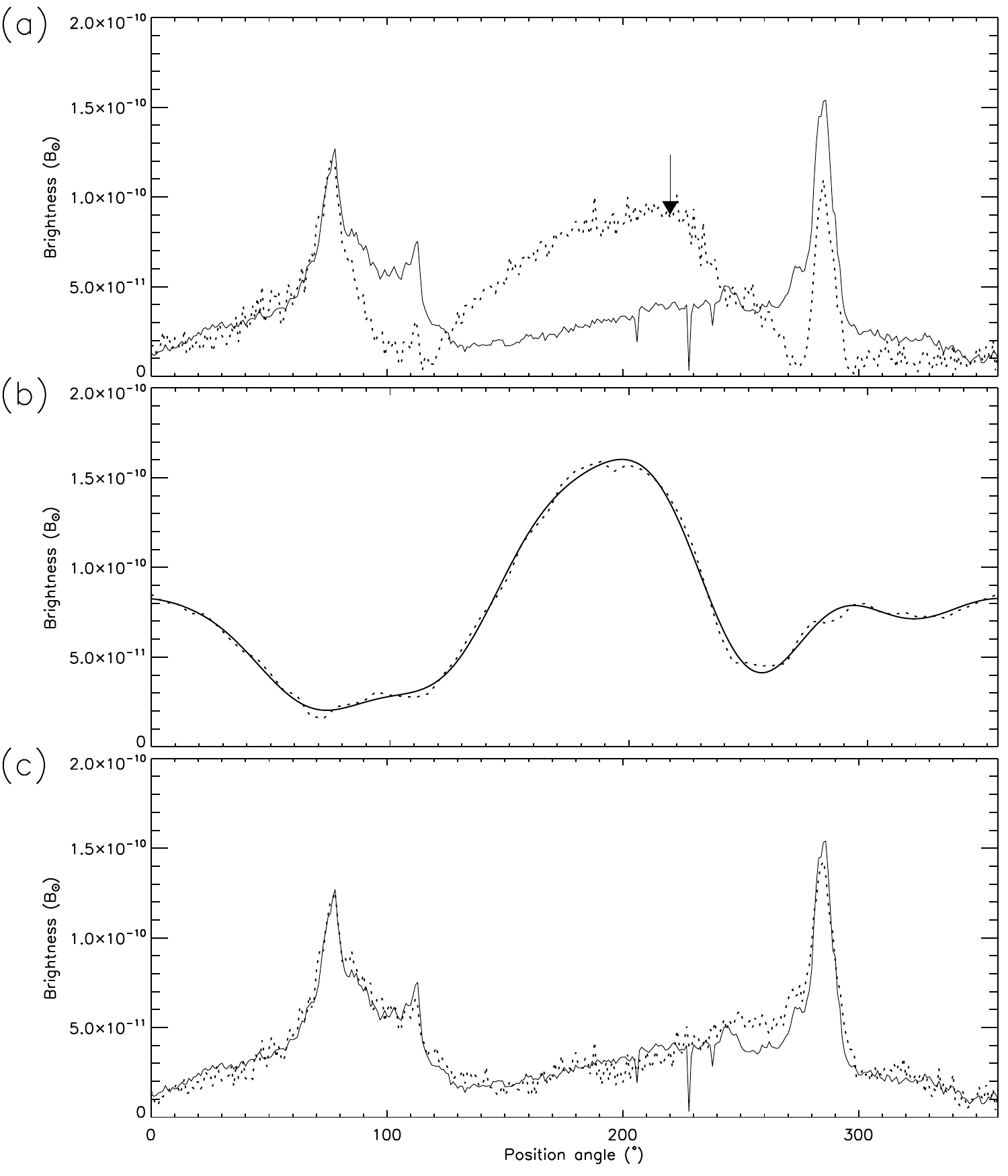}
\end{center}
\caption{(a) Comparison of SECCHI COR2 A (solid line) and B (dotted) polarized brightness at a height of 5.0\Rs\ for observations made at 2007/03/18 02:27. The arrow points to the peak of a broad feature of stray light contamination. (b) Subtraction profile for COR2 B (see text). (c) As (a), but with correction factors applied to COR2 B and application of the subtraction factor.}
\label{corab}
\end{figure}

Figure \ref{corab}a shows the polarized brightness profile of both coronagraphs at a height of 5.0\Rs\ and date 2007/03/18 02:27. The peak of the stray light contamination is indicated by the arrow, and the agreement between the two coronagraphs is poor. For a given height, we compare the COR2A \& B \pB\ values over the period 2007/03/16-2007/03/29. The fitting procedure searches three different corrective factors $f_1, f_2, f_3$ for the three individual polarizer images for COR2B (at 0, 120 and 240\de\ respectively). The initial value for each factor is 1. For each combination of corrective factors, a subtraction profile is calculated as the difference between COR2A and B polarized brightness averaged over the two week period. This is calculated for each position angle bin. The average difference is then subtracted from the COR2B profile, and a score assigned to the fit as the mean absolute percentage deviation. A search is made for the best fit by minimising the score, with $f_1, f_2, f_3$ the search parameters. At a height of 5.0\Rs, the best fit is found for $f_1=0.83, f_2=0.78, f_3=0.76$, for which the subtraction profile is shown in figure \ref{corab}b. For convenience, this subtraction profile is fitted to a truncated $6^{th}$-order sine series as a series of position angle $\Omega$, overplotted in figure \ref{corab}b. The sine series is given by

\begin{equation}
S (\Omega) = p + \sum\limits_{k=1}^6 \left(\alpha_k \sin(k\Omega) + \beta_k \cos(k\Omega) \right).
\label{subfact}
\end{equation} 

\begin{figure}[]
\begin{center}
\includegraphics[width=8.0cm]{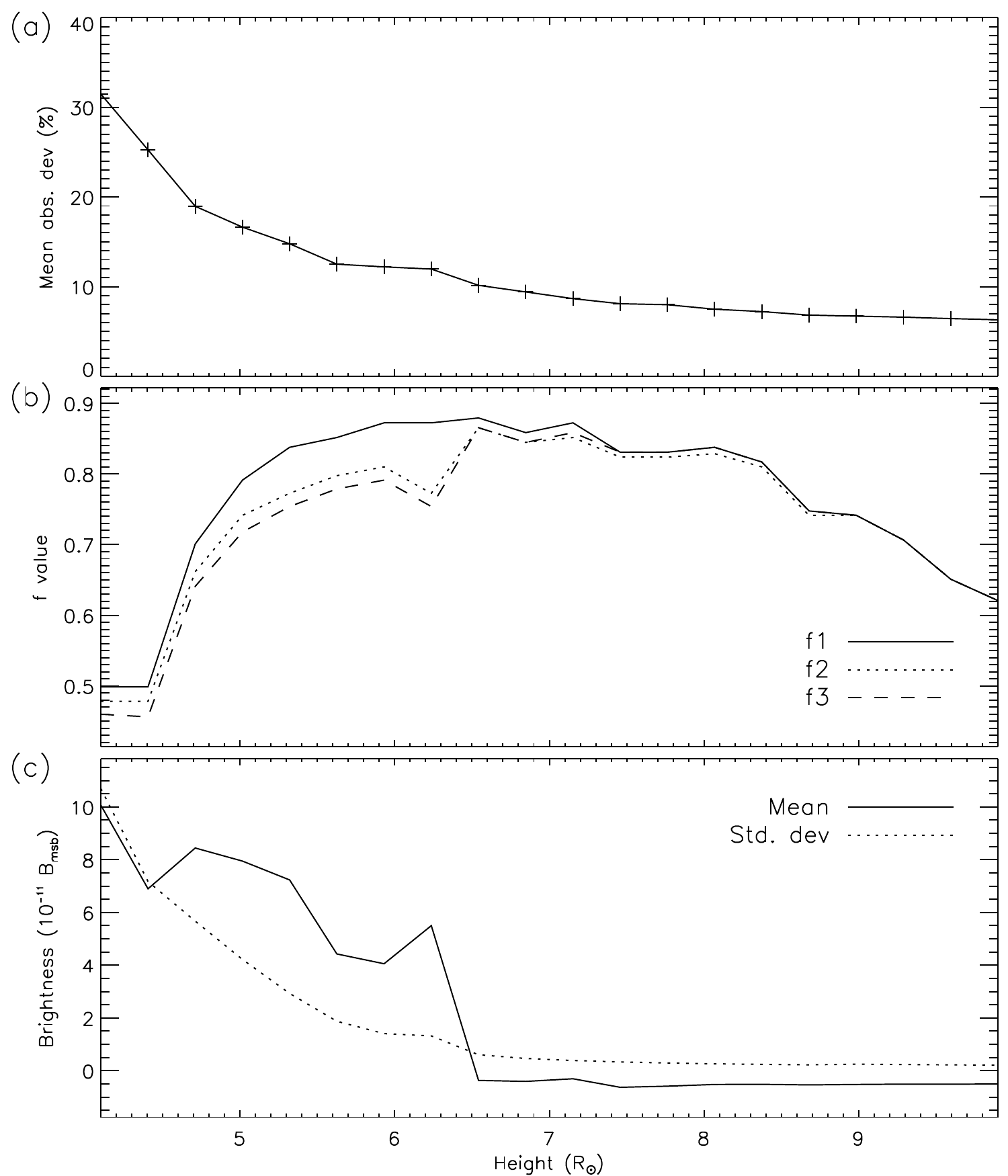}
\end{center}
\caption{(a) Mean absolute deviation of COR2 A and B \pB\ observations as a function of height, following the correction procedures described in the text. The mean absolute deviation is calculated for all \pB\ observations between 2007/03/16-29. (b) Values for $f_{1,2,3}$ as listed in table \ref{corab_table}. (c) The mean and standard deviation across all position angles of the COR2 B subtraction profiles, created using equation \ref{subfact} and the parameters of table \ref{corab_table}. This shows how the amplitude and mean of the subtraction profile generally decreases with height.}
\label{corab2}
\end{figure}

The corrected profile is shown in figure \ref{corab}c. The agreement with COR2A is much improved, with a mean absolute deviation of $\sim$16\%\ over the whole time period. This fitting approach can be repeated for many heights. The stray light feature becomes relatively less bright with increasing height, and the agreement between COR2A and B improves with height. The correction factors also take values closer to 1 with increasing height. Values for $f_{1,2,3}$, $p$, $\alpha_k$ and $\beta_k$ are listed in table \ref{corab_table} for several heights between 4 and 5.9\Rs. Figure \ref{corab2}a shows how the fit between COR2 A and B improves with height. Figure \ref{corab2}b shows the changing values of $f_1, f_2, f_3$ with height, and figure \ref{corab2}c shows how the mean and standard deviation of the subtraction profiles (calculated over all position angles) tend to decrease with height.

\begin{table}[h]
\tiny
\begin{center}
\begin{tabular}{ c | c | c | c | c | c | c | c | c | c | c | c | c | c | c }
{\Rs} & {$f_1$, $f_2$, $f_3$} & {$p$} & {$\alpha_1$} & {$\beta_1$} & {$\alpha_2$} & {$\beta_2$} & {$\alpha_3$} & {$\beta_3$} & {$\alpha_4$} & {$\beta_4$} & {$\alpha_5$} & {$\beta_5$} & {$\alpha_6$} & {$\beta_6$} \\
\hline
4.10 & 0.50, 0.48, 0.46 & 100.73 & -46.80 & -64.70 & 14.38 & 124.50 & -13.07 & -1.23 & 18.29 & -9.69 & 7.07 & 3.99 & -7.84 & -6.65 \\
4.41 & 0.50, 0.48, 0.46 & 68.97 & -34.24 & -41.54 & 9.13 & 82.64 & -9.38 & -2.74 & 14.87 & -6.34 & 3.68 & 3.94 & -7.65 & -3.00 \\
4.71 & 0.70, 0.66, 0.64 & 84.44 & -29.13 & -36.16 & 3.71 & 61.84 & -7.26 & -10.59 & 13.59 & 0.07 & 1.99 & 5.54 & -5.13 & -2.93 \\
5.02 & 0.79, 0.74, 0.72 & 79.48 & -24.21 & -28.57 & 3.14 & 42.74 & -5.32 & -11.06 & 10.55 & 2.31 & 1.24 & 4.58 & -4.01 & -2.44 \\
5.32 & 0.84, 0.77, 0.75 & 72.34 & -19.53 & -20.15 & -0.10 & 26.63 & -2.30 & -10.07 & 8.19 & 4.35 & 0.26 & 3.34 & -2.99 & -1.23 \\
5.63 & 0.85, 0.80, 0.78 & 44.28 & -11.30 & -12.91 & -0.85 & 16.49 & -2.21 & -8.24 & 6.01 & 3.37 & -0.41 & 2.73 & -2.20 & -0.71 \\
5.93 & 0.87, 0.81, 0.79 & 40.57 & -9.55 & -10.53 & -1.29 & 9.50 & -1.47 & -7.85 & 4.47 & 3.22 & -0.44 & 1.79 & -1.92 & -0.28 \\
6.24 & 0.87, 0.77, 0.75 & 55.01 & -12.33 & -9.80 & -1.40 & 3.21 & -0.51 & -6.88 & 4.33 & 4.03 & -0.27 & 1.15 & -1.56 & -0.36 \\
6.54 & 0.88, 0.87, 0.87 & -3.69 & 2.09 & -2.66 & -0.57 & 3.90 & -1.03 & -6.13 & 1.56 & -0.03 & -1.26 & 1.25 & -1.20 & -0.35 \\
6.85 & 0.86, 0.84, 0.84 & -4.06 & 1.78 & -1.43 & -1.10 & 2.15 & -0.61 & -5.21 & 1.27 & 0.10 & -1.18 & 0.75 & -0.87 & -0.01 \\
7.15 & 0.87, 0.85, 0.86 & -3.05 & 2.07 & -0.66 & -0.75 & 0.64 & 0.74 & -4.62 & 0.87 & 0.76 & -0.78 & 0.41 & -0.30 & 0.07 \\
7.46 & 0.83, 0.82, 0.83 & -6.29 & 2.12 & 1.94 & -0.99 & 2.22 & 0.28 & -2.54 & 0.16 & 0.22 & -0.92 & 0.46 & -0.54 & 0.45 \\
7.76 & 0.83, 0.82, 0.83 & -5.87 & 1.82 & 2.08 & -1.13 & 1.62 & 0.27 & -2.03 & 0.18 & 0.19 & -0.84 & 0.28 & -0.41 & 0.54 \\
8.07 & 0.84, 0.83, 0.84 & -5.22 & 1.65 & 2.18 & -0.95 & 0.92 & 0.70 & -1.70 & -0.06 & 0.43 & -0.64 & 0.03 & -0.22 & 0.60 \\
8.37 & 0.82, 0.81, 0.82 & -5.17 & 1.30 & 2.24 & -1.30 & 0.72 & 0.52 & -1.23 & 0.09 & 0.04 & -0.68 & -0.02 & -0.17 & 0.54 \\
8.68 & 0.75, 0.74, 0.75 & -5.39 & 1.00 & 2.04 & -1.40 & 1.00 & 0.49 & -0.84 & 0.12 & -0.24 & -0.53 & -0.05 & -0.17 & 0.58 \\
8.98 & 0.74, 0.74, 0.74 & -5.20 & 0.75 & 2.09 & -2.18 & 1.13 & 0.12 & -0.47 & 0.33 & -0.70 & -0.61 & 0.22 & -0.10 & 0.42 \\
9.29 & 0.71, 0.71, 0.71 & -5.08 & 0.61 & 1.91 & -2.07 & 1.14 & 0.24 & -0.30 & 0.30 & -0.76 & -0.46 & 0.12 & -0.13 & 0.40 \\
9.59 & 0.65, 0.65, 0.65 & -5.11 & 0.48 & 1.67 & -1.96 & 1.35 & 0.27 & -0.23 & 0.25 & -0.84 & -0.35 & 0.06 & -0.12 & 0.40 \\
9.90 & 0.62, 0.62, 0.62 & -5.00 & 0.36 & 1.45 & -1.88 & 1.43 & 0.24 & -0.10 & 0.21 & -0.87 & -0.28 & 0.03 & -0.12 & 0.44 \\
\end{tabular} 
\caption{Correction factors for the COR2 B polarizer observations. Values for $p$, $\alpha_k$ and $\beta_k$ are in units of $10^{-12}$\msb.}
\label{corab_table}
\end{center}
\end{table}

In summary, this section provides a method which forces the COR2 B \pB\ observations to closely match those of COR2 A. To correct COR2 B at a given height and position angle, the individual polarizer images are multiplied by the appropriate $f_{1,2,3}$ factors, and a value is subtracted according to equation \ref{subfact} and the parameters of table \ref{corab_table}. The good agreement (e.g. figure \ref{corab}c) is not surprising given the creation of the subtraction factor. This however is the major flaw in the method - it is not based on a simple adjustment or a single calibration factor for each polarizer angle. There are different calibration factors at different heights, as well as a complicated subtraction factor which also differs with height. It is difficult to determine whether the problems associated with the COR2 B instrument are stable over time, and whether our solution is valid over long time periods. \paperii\ will address this issue by comparing tomography results from the three coronagraphs at different times during their missions.

\section{Application of calibrated data \\and future work}
\label{application}

\subsection{Tomography}
The primary purpose of developing the DST and calibration methods is for use in coronal rotation tomography. In \paperii, calibrated data will be used in a tomography method to gain maps of the coronal electron density at a height of 5.0\Rs. This height is chosen as an optimal height for the tomography method and as a height where effective crosscalibration may be applied. The important parts of the processing and calibration methods presented here are the reduction of signal of large CMEs by the DST, the conversion of LASCO C2 total brightness observations into an approximation of \Bk, and the crosscalibration between coronagraphs. These are crucial steps in achieving reasonable reconstructions of the coronal density. 

\begin{figure}[h]
\begin{center}
\includegraphics[width=8.0cm]{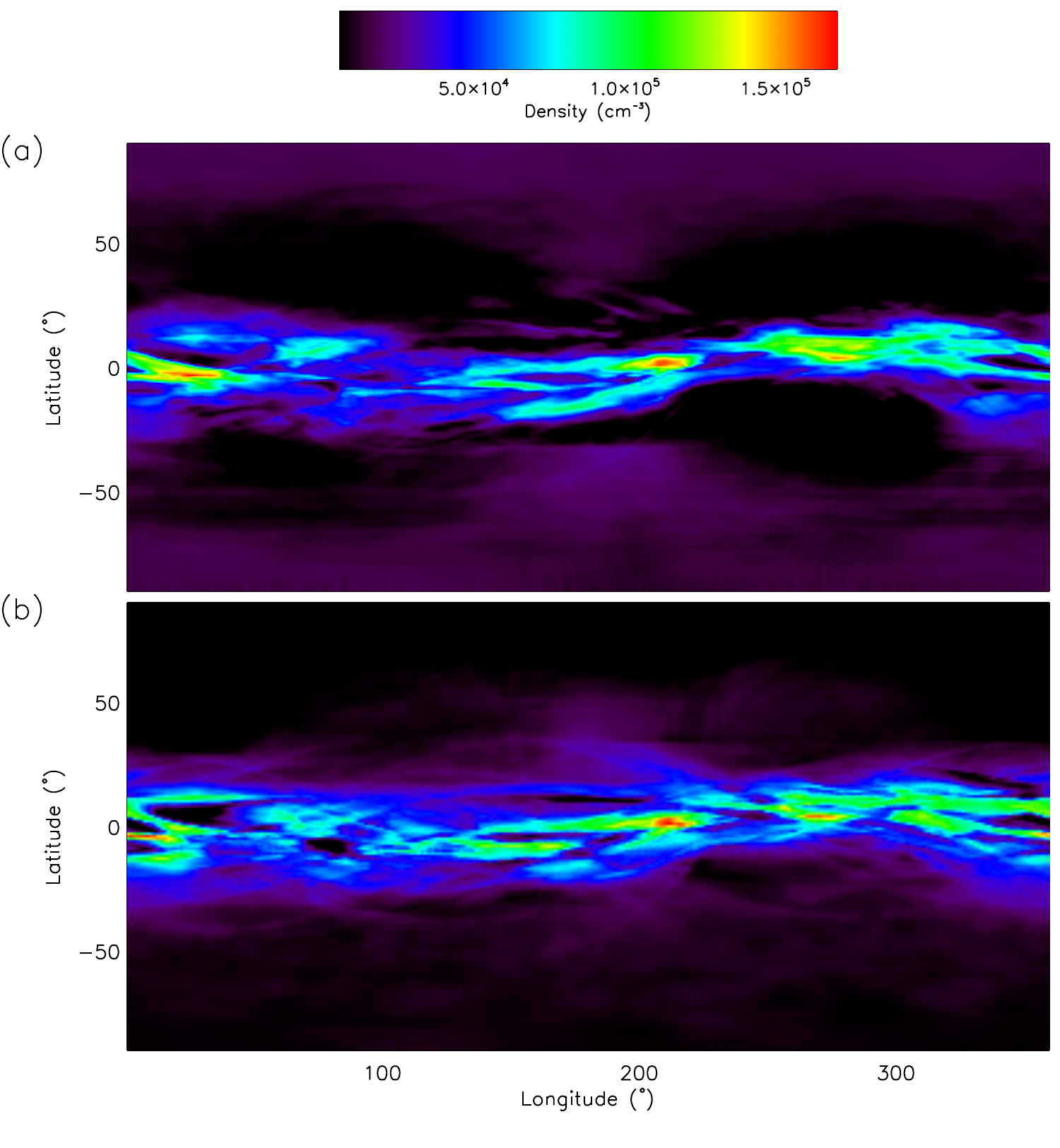}
\end{center}
\caption{Tomographical reconstructions of the coronal electron density for a spherical shell at a height of 5.0\Rs\ for (a) LASCO C2 and (b) COR2 A. Axis show Carrington longitude and latitude.}
\label{tomo1}
\end{figure}

In anticipation of \paperii, figure \ref{tomo1} show two tomography maps of the coronal electron density, at a height of 5.0\Rs, created from LASCO C2 and COR2A data between 2007/03/15-30. The structural agreement between the two is excellent. There are regions of unphysically low density within and neighbouring the streamer belts in both maps. This has always been a problem of coronal rotational tomography. It is likely caused both by time variation of the high-density, bright streamers and an artifact of the method itself and will be addressed in \paperii. The LASCO C2 reconstruction appears smoother and cleaner because a larger number of observations are available for the reconstruction. 

\begin{figure}[]
\begin{center}
\includegraphics[width=8.0cm]{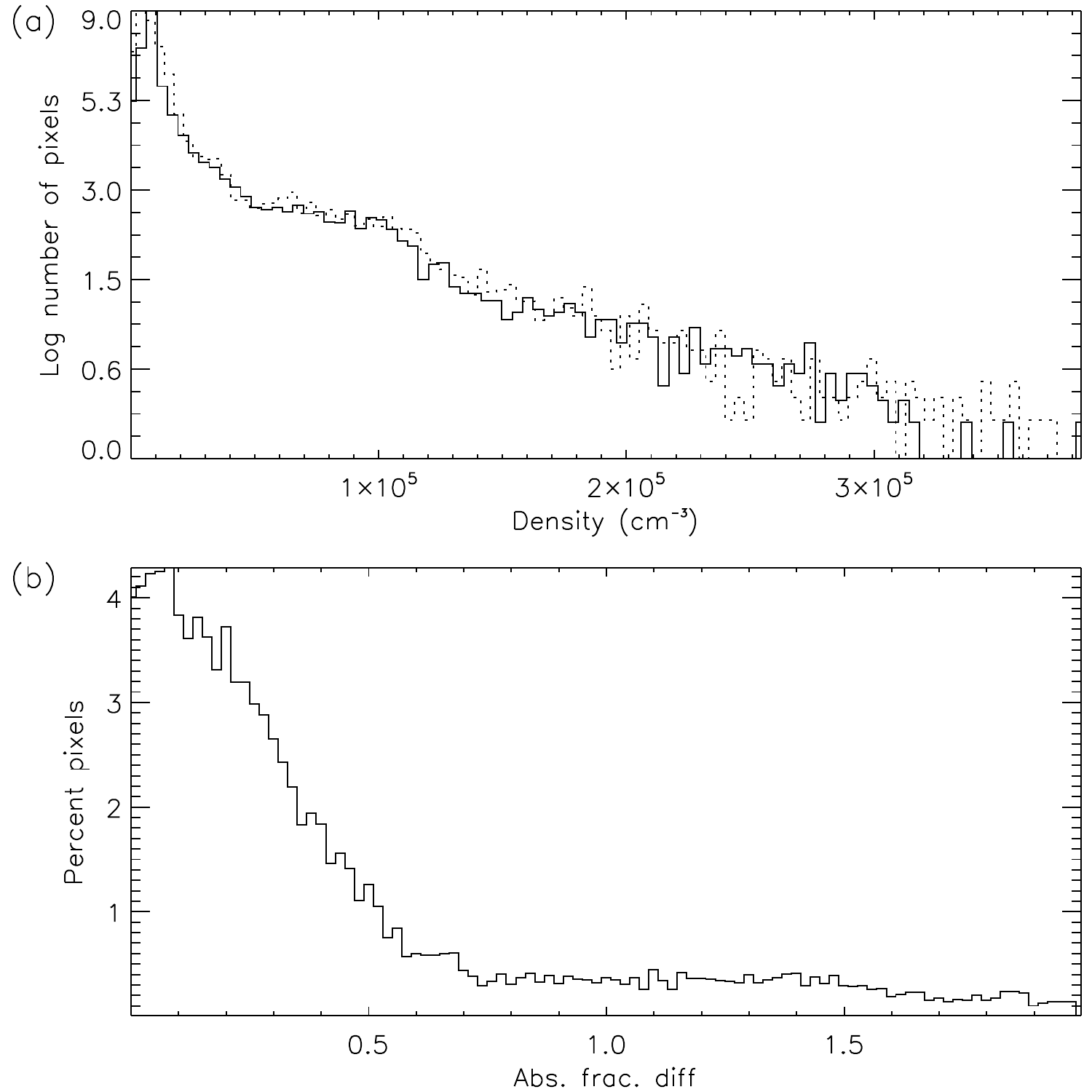}
\end{center}
\caption{(a) Histogram showing the distribution of densities within the tomography maps for COR2 A (solid line) and LASCO C2 (dotted). (b) Histogram showing distribution of absolute fractional differences between the two maps.}
\label{tomo2}
\end{figure}

The comparison of tomography maps made by different coronagraphs enables an interesting test of the reliability of the tomography. There is reasonable numerical agreement between the reconstructed densities, as shown in figure \ref{tomo2}a. The most probable density (corresponding to the large polar coronal holes) is $\sim10^4$\vcm, and the streamers are $\sim10^5$\vcm, agreeing well with values found by others \citep{doyle1999, guhathakurta1996, gibson2003}. The distribution of densities agree well between the two coronagraphs, which is to be expected given the good agreement in brightness following the calibration method. Figure \ref{tomo2}b shows the distribution of absolute fractional differences over all map pixels. The agreement is reasonable with 67\%\ of pixels agreeing within 38\%\ or less. 

Combining the data from two or three coronagraphs viewing the corona from different angles enables a reconstruction to be made using a shorter time period of observations. This enables an improved reconstruction because any changes in the coronal structure will have a lesser impact. It also helps in analysing temporal changes in coronal structure. The brief results shown here show promise that such an approach is possible. A full description of the tomography method and results will be presented in \paperii\ and \paperiii. 

\subsection{CME masses and other diagnostics}
\begin{figure}[]
\begin{center}
\includegraphics[width=8.0cm]{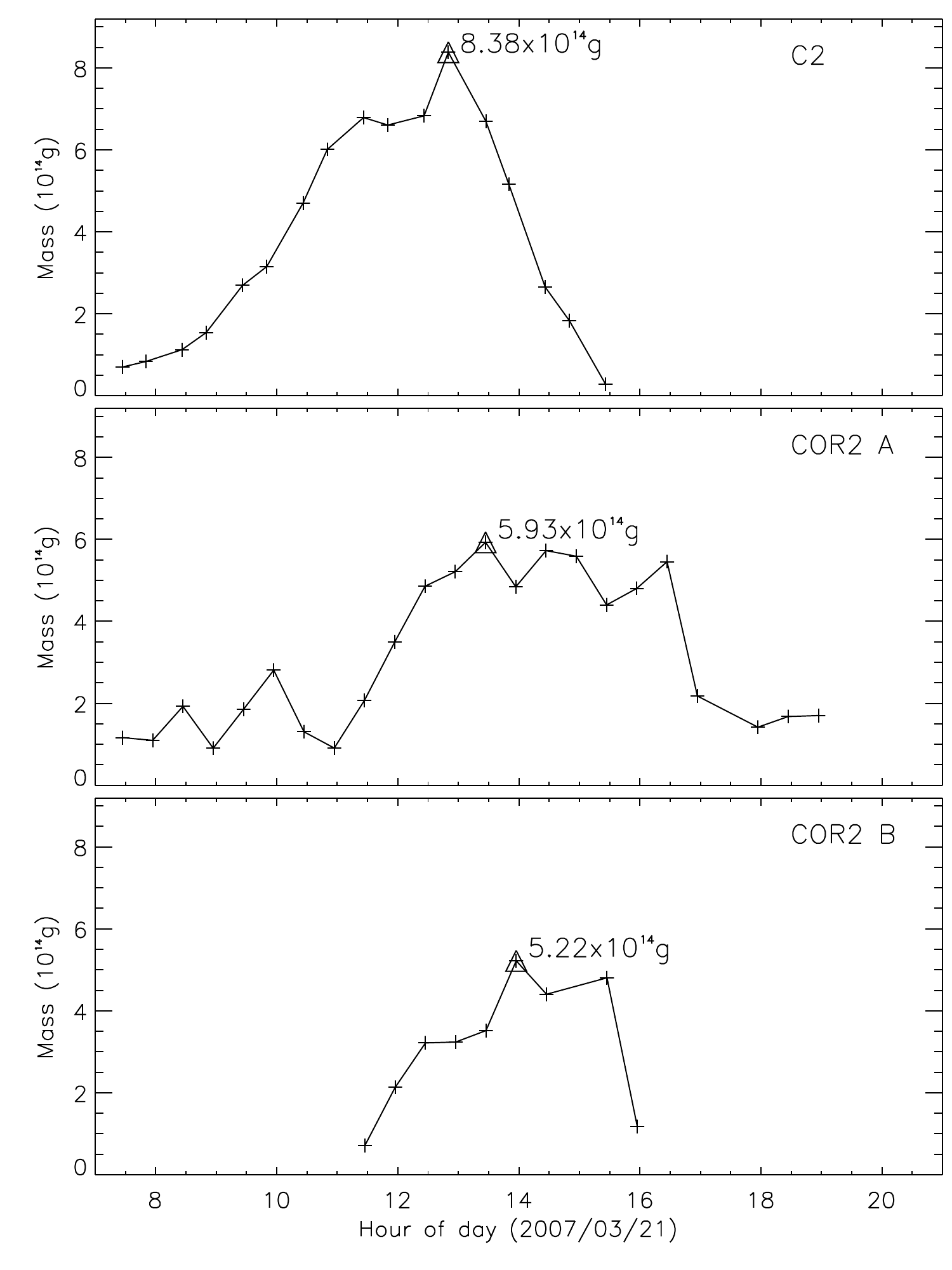}
\end{center}
\caption{Estimated CME mass as function of time for 2007/03/21, integrated over position angles 20-140\de\ for the three coronagraphs, as labelled in the panels. The maximum estimated mass (i.e. the best estimate for the CME mass) is labelled and indicated by a triangle. The plane-of-sky assumption is used.}
\label{cme1}
\end{figure}

The calibrated data may be used to gain useful 3D information on the spatial distribution of CMEs, and an estimate of their mass. The DST processing is a crucial part of this study because it allows study of the CME in absence of the background structures, and without the difficulty of interpreting time-differenced images. Figure \ref{cme1} shows the estimated mass as a function of time for the north-east CME of 2007/03/21 shown earlier in the context of DST processing. For calibrating the dynamic component DST images, there is no background subtraction since the static background has already been removed. The only calibrating factors therefore are the multiplicative ones. Assuming that the emitting electrons are solely in the plane of sky, the masses are gained by simple inversion. 

The best estimate found for the CME mass in LASCO C2 is $8.38\times10^{14}$g. The masses found by the COR2 A\&B coronagraphs are 5.93 and $5.22\times10^{14}$g respectively. There are several reasons for this difference. The used LASCO C2 field of view is restricted compared to the COR2s, although if the field of view is smaller we would expect the mass estimate to be lower rather than higher. Another, more physical, and potentially useful source of the difference is the true spatial distribution of the CME. The fact that the three profiles peak at slightly different times suggest that the CME is not in the plane of sky. The LASCO C2 profile is expected to be shifted due to the different height range, but there is also a small shift between the COR2. In principle, this shift and the different estimated masses may be used to constrain the true longitudinal position of the CME. This will be studied in a future work.

The most worrying discrepancy is between the Coordinated Data Analysis Workshop (CDAW, \citet{yashiro2004,vourlidas2002}) CME catalog estimate of mass, $2.4\times10^{15}$g, which is almost three times our estimate. The standard estimate is based on identifying the region of interest in time-differenced C3 images. The inversion method steps are identical (i.e. plane-of-sky approximation, and identical forms of the Thomson-scattering electron emission model). The difference should not be so high because the whole CME in this example is contained within the LASCO C2 field of view at the time of peak mass (i.e. the CME is not partially measured due to being too large to fit in the field of view). We believe one reason is the movement of background streamers in reaction to the passing of a CME. Note in figure \ref{figsep2}b how the existing background streamer increases in width due to the passage of the CME. Often with large CMEs, the streamer will split into two for a short period. This change is not contaminating the separated dynamic image of \ref{figsep2}c therefore is not included in the CME mass estimate. Time-differencing, however, would include changes in the configuration of the background corona in the estimate of CME mass. This, and other aspects, will be investigated in a future work.

\subsection{F-corona diagnostics}

A potential application of the calibration methods will be the use of the calibrated backgrounds to estimate the F-corona brightness. The F-corona brightness \Bf\ gives information on the composition and spatial distribution of interplanetary dust, and the influence of the Sun on the dust \citep[e.g.,][]{mann1992}.  A comprehensive review of interplanetary dust and the F corona can be found in \citet{grun2001}, and references within. As seen in figure \ref{testlt}, there are instrumental features in the background which may be identified and removed by taking advantage of the regular spacecraft roll manoevers. This would leave a smooth background which is probably dominated by the F-corona brightness. This part of the signal is interesting as an unique measurement of the dust emission close to the Sun. A former study has shown that the dust emission changes little if at all between solar minimum and maximum \citep{morgan2007fcorona}, with implications for the interaction of the plasma with dust. Variations in the F-corona over smaller timescales (days or weeks) is more difficult to interpret because any rapid changes are probably suspect - they would probably be due to contamination by \Bk. We are interested in processing and studying the LASCO C2 background images in order to extend the temporal study of \citet{morgan2007fcorona}, and to study the large-scale structure of the dust, possibly using tomography and a model of dust emission.

\section{Summary}
\label{summary}

This study focuses on heights of overlap of the SECCHI COR2 and LASCO C2 coronagraphs, namely $\sim$4.1-6.0\Rs, and focuses closely on a height of 5.0\Rs\ as this will be used as a basis for tomography in \paperii. The paper also concentrates on a period of a few weeks in 2007/03. At this time, the three spacecraft were close together and making regular observations, and the corona was in a quiet state, giving an ideal time for crosscalibration.

The main methods presented are:
\begin{itemize}
\item DST which uses a spatial-temporal deconvolution method to separate dynamic events from quiescent background.
\item Calibration of LASCO C2 \pB\ observations using stars. The new calibration factors result in the elimination of strange artifacts such as the dip in brightness towards the equator, most apparent during solar minimum configuration.
\item Conversion of LASCO C2 \pB\ observation sequences into \Bk\ images through inversion and reintegration.
\item Conversion of total brightness observations (without polarization) into \Bk\ by subtraction of a long-term median background created from comparing the total brightness observations with the \Bk\ images (LASCO C2 only). The long-term backgrounds are shown to be very stable with the largest oscillations in brightness with periods matching the SOHO orbit, and a smaller oscillation linked to the coronal rotation. Following this conversion, the \Bk\ images gained from the highest-cadence total brightness images agree very well with the \Bk\ images gained directly from the lower-cadence \pB\ observations.
\item Crosscalibration of COR2 A to match LASCO C2 by application of simple calibration factors. This simple correction results in a huge improvement in agreement between the two coronagraphs in a limited height range close to 5\Rs.
\item Crosscalibration of COR2 B to match COR2 A by functions of position angles gained by empirical determination of calibration factors and a subtraction background (to remove a prominent stray light feature in COR2 B). The crosscalibration is good for the time period 2007/03, and \paperii\ will use tomography results to establish whether the cross-calibration is dependable at other times.
\end{itemize}

An important assumption made is the use of LASCO C2 as a standard, against which the other coronagraphs are adjusted. This is a choice based on the accuracy of calibration of LASCO C2 data, based on analysis of stars to test the vignetting correction and correct the \pB\ sequence calibration (this work) and absolute radiometric calibration of total brightness observations described by \citet{colaninno2015}. The main weakness is the limited field of view where the cross-calibration method is reliable. Ideally, similar methods should be applied to the LASCO C3 instrument. This would provide a far greater overlap of field of view with SECCHI COR2. This is more difficult than the methods presented here for LASCO C2 since the F-corona becomes increasingly polarized at heights above $\sim$6\Rs, and is more difficult in practice due to lower signal to noise and spatial resolution.

In \paperii\ and \iii, the calibrated and processed data will be used for tomography. Other future work will focus on CME and F-corona diagnostics, as well as extensions to the calibration methodology. In particular, it would be valuable to extend the star calibration to the COR2 A \& B coronagraphs. This is difficult in practice due to the higher signal-to-noise of these instruments compared to LASCO C2.

 \begin{acknowledgements}
Huw is grateful for funding from the Coleg Cymraeg Cenedlaethol, and of the Leverhulme Trust to Prifysgol Aberystwyth. Without their support this work would not be possible. I am also grateful for the comments of an anonymous referee which helped me greatly in improving the methodology. The SOHO/LASCO data used here are produced by a consortium of the Naval Research Laboratory (USA), Max-Planck-Institut fuer Aeronomie (Germany)), Laboratoire d'Astronomie (France), and the University of Birmingham (UK). SOHO is a project of international cooperation between ESA and NASA. The STEREO/SECCHI project is an international consortium of the Naval Research Laboratory (USA), Lockheed Martin Solar and Astrophysics Lab (USA), NASA Goddard Space Flight Center (USA), Rutherford Appleton Laboratory (UK), University of Birmingham (UK), Max-Planck-Institut fu\"r Sonnen-systemforschung (Germany), Centre Spatial de Liege (Belgium), Institut d�Optique Th\'eorique et Appliq\'uee (France), and Institut d'Astrophysique Spatiale (France). 
 \end{acknowledgements}

 \clearpage

 \clearpage

 \clearpage

 \clearpage

 \clearpage

 \clearpage

 \clearpage

 \clearpage

 \clearpage
\begin{figure}[]
\begin{center}
\includegraphics[width=12.0cm]{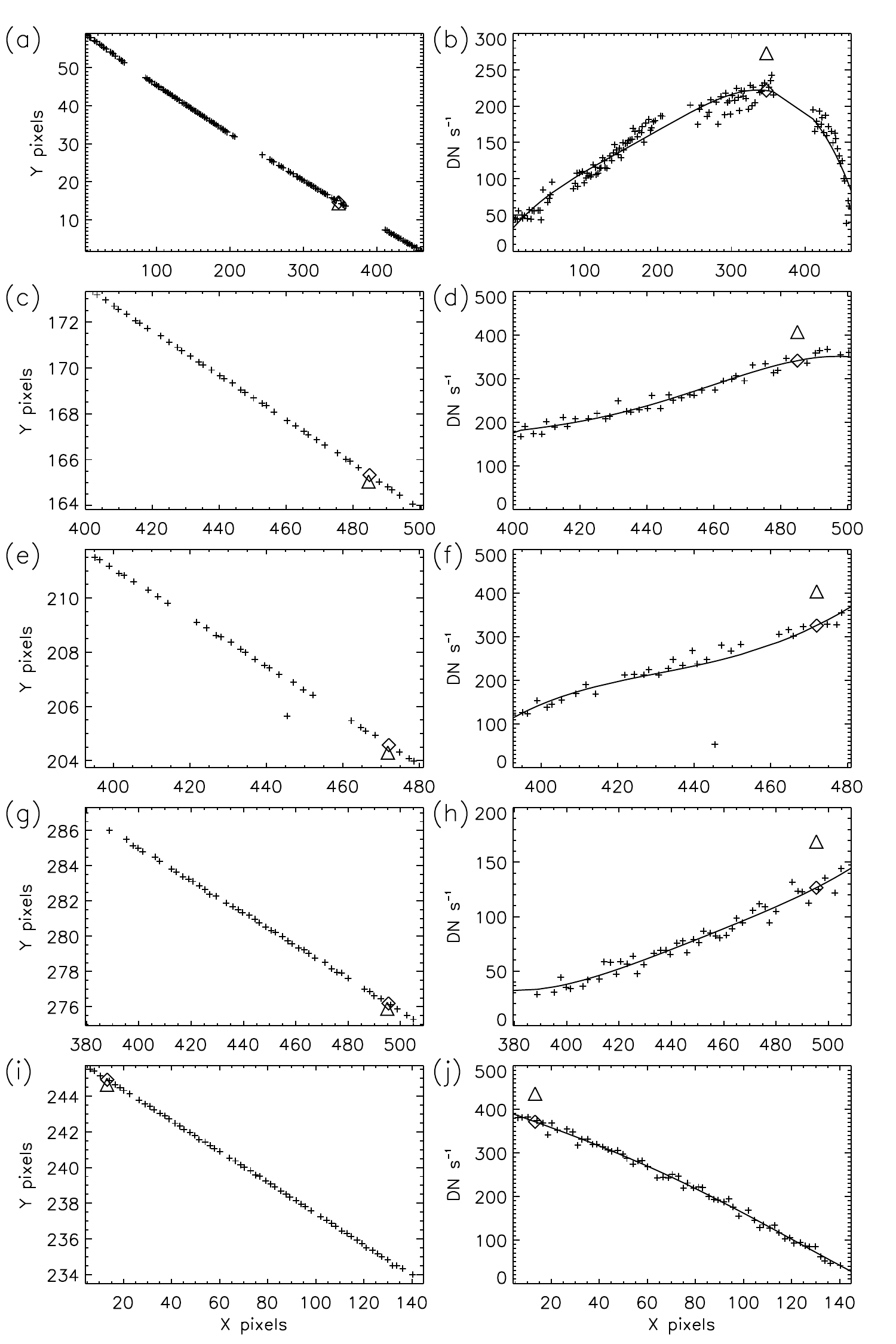}
\end{center}
\caption{Left column - plots of several LASCO C2 star tracks across image $X-Y$ for \Btot\ (crosses) and the corresponding star detected in \pB\ images (triangles). Fitting a straight line to the \Btot\ star track enables interpolation to the time of the \pB\ observation, giving the expected position of the star. The expected position is given by the diamond. Right column - the measured brightness of each star for \Btot\ (crosses) and \pB\ (triangle). A $4^{th}$-order polynomial is fitted to the \Btot\ brightness, giving the expected brightness of the star at the time of the \pB\ observation (diamond). The results here are shown for the clear polarizer state.}
\label{pbratiocalc}
\end{figure}

 \clearpage

 \clearpage

 \clearpage

 \clearpage

 \clearpage

 \clearpage

 \clearpage

 \clearpage

 \clearpage

 \clearpage

 \clearpage
\begin{figure}[]
\begin{center}
\includegraphics[width=8.0cm]{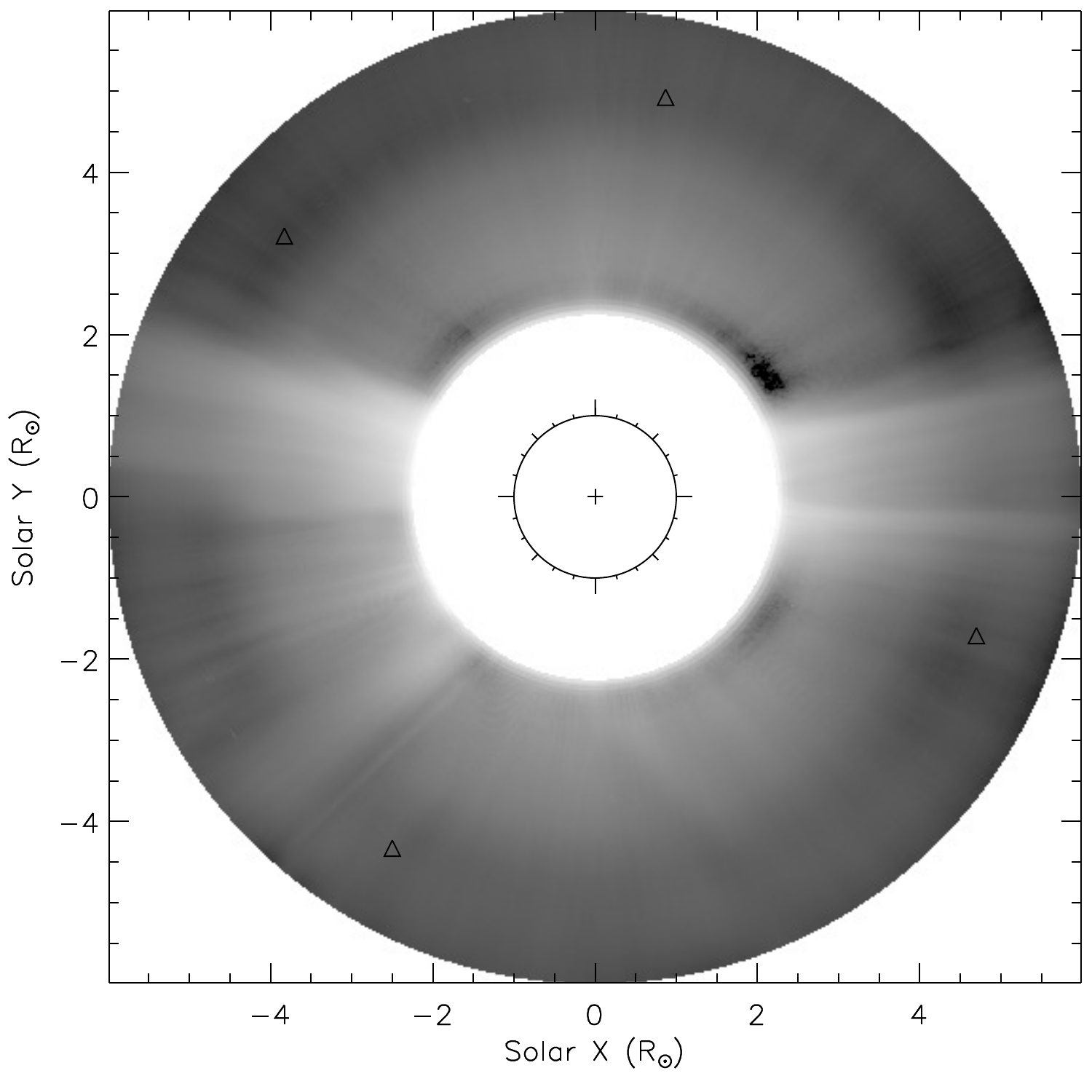}
\end{center}
\caption{Calibrated, background-subtracted quiescent component total brightness image of 2007/03/21 11:26 (original image shown without DST processing and calibration in figure \ref{figsep}a). Log$_{10}$ brightness is shown. The four triangles are relevant to figure \ref{wlcal}.}
\label{calapp}
\end{figure}

 \clearpage
\begin{figure}[]
\begin{center}
\includegraphics[width=8.0cm]{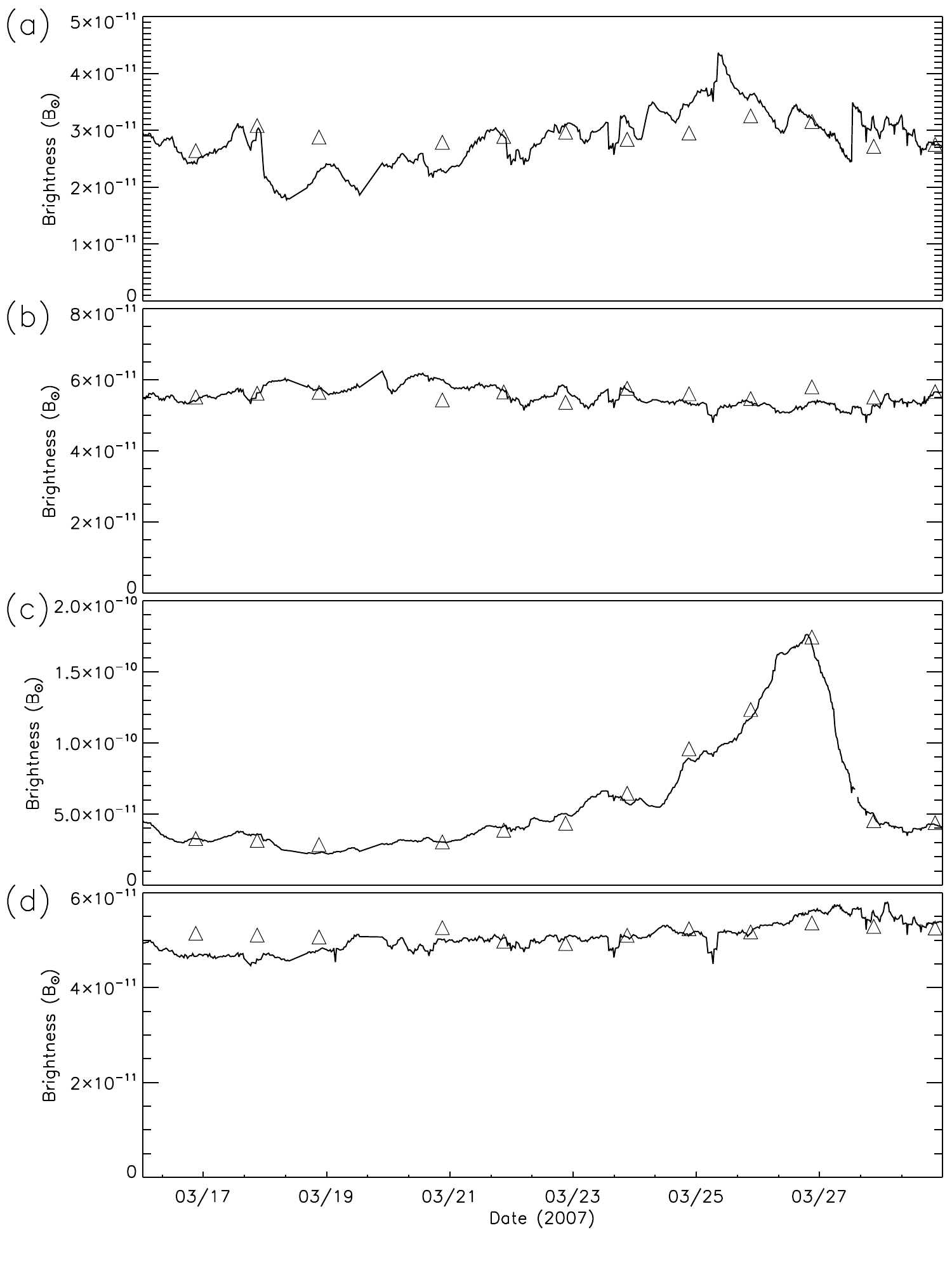}
\end{center}
\caption{Comparison of calibrated total brightness values with \Bk\ values over time at a height of 5.0\Rs\ and position angles (a) 50\de, (b) 150\de, (c) 250\de, and (d) 350\de\ (at the position of the triangles in figure \ref{calapp}). The lines give the calibrated total brightness values and the triangles give the \Bk\ values.}
\label{wlcal}
\end{figure}


 \clearpage

 \clearpage

 \clearpage

 \clearpage


 \clearpage

 \clearpage

 \clearpage

\end{document}